\documentclass[12pt]{article}
\pdfoutput=1
\usepackage{jheppub}

\newcommand{\CN}{{\cal N}}

\newcommand{\CM}{{\cal M}}

\newcommand{\CC}{{C}}
\newcommand{\CZ}{{Z}}
\newcommand{\CH}{{\cal H}}

\newcommand{\C}{{\mathbb C}}
\newcommand{\bA}{{\mathbb A}}

\DeclareMathOperator{\Bun}{Bun}
\DeclareMathOperator{\Loc}{Loc}
\DeclareMathOperator{\Sb}{Sb}
\DeclareMathOperator{\Fc}{Fc}
\DeclareMathOperator{\Ai}{Ai}

\title{Twisted compactifications of 3d $\CN=4$ theories and conformal blocks}
\abstract{Three-dimensional $\CN=4$ supersymmetric quantum field theories admit two topological twists, the Rozansky-Witten twist and its mirror. 
Either twist can be used to define a supersymmetric compactification on a Riemann surface and a corresponding space of supersymmetric ground states. 
These spaces of ground states can play an interesting role in the Geometric Langlands program.
We propose a description of these spaces as conformal blocks for certain non-unitary Vertex Operator Algebras
and test our conjecture in some important examples. The two VOAs can be constructed respectively from a UV Lagrangian description of the $\CN=4$ theory 
or of its mirror. We further conjecture that the VOAs associated to an $\CN=4$ SQFT inherit properties of the theory which only emerge in the IR, such as 
enhanced global symmetries. Thus knowledge of the VOAs should allow one to compute the spaces of supersymmetric ground states 
for a theory coupled to supersymmetric background connections for the full symmetry group of the IR SCFT. 
In particular, we propose a conformal field theory description of the spaces of ground states for the $T[SU(N)]$ theories. 
These theories play a role of S-duality kernel in maximally supersymmetric $SU(N)$ gauge theory and thus the corresponding spaces of supersymmetric ground states should provide a kernel for the Geometric Langlands duality for special unitary groups.}

\author[1]{Davide Gaiotto}
\affiliation[1]{Perimeter Institute for Theoretical Physics, Waterloo, Ontario, Canada N2L 2Y5}

\begin{document}
\maketitle
\section{Introduction}
This is a companion paper to \cite{Gaiotto:2016aa}. The main subject of this paper are $\CN=4$ SQFT in three dimensions, equipped with interesting global symmetry groups.
Such three-dimensional SQFTs can appear as boundary degrees of freedom for half-BPS boundary conditions in $\CN=4$ Super Yang Mills theory \cite{Gaiotto:2008sa}. 

In turns, half-BPS boundary conditions in $\CN=4$ SYM descend, upon compactification on a Riemann surface $C$, to branes which play an important role in the 
gauge theory interpretation of the Geometric Langlands duality \cite{Kapustin:2006pk,Witten:2008ep,Witten:2009mh,Witten:2015aa}. 
When the original boundary conditions involve boundary degrees of freedom in the form of 3d $\CN=4$ SQFT, 
the Chan-Paton bundles for the corresponding branes arise as spaces of super-symmetric 
ground states on $C$ for the corresponding 3d SQFTs \cite{Gaiotto:2016aa}. 

The objective of this paper is to find a characterization of these Chan-Paton bundles which is flexible enough to overcome a crucial problem: 
the $\CN=4$ SQFTs we are interested in often have important low energy symmetry groups which are not fully visible in any known UV Lagrangian descriptions.

Our strategy is to associate to each $\CN=4$ SQFT a Vertex Operator Algebra whose conformal blocks on $C$ match the desired Chan-Paton spaces 
and whose symmetries match the low energy symmetries of the SQFT. 

Some elements of our proposal are somewhat conjectural and the Vertex Operator Algebra which occur in practice in our calculations 
are rather intricate and non-unitary. The main focus of this paper is to analyze some important basic examples 
and to collect evidence that the VOA associated to linear quiver gauge theories of unitary groups have hidden symmetries which 
match the known IR symmetry enhancement of the gauge theories. 

For the very simplest examples, we will also describe in some detail the interpretation of conformal blocks as Chan-Paton bundles
for branes. The definition of conformal blocks for the non-unitary VOA which occur in our setup have subtleties which may be somewhat unfamiliar to physicists 
(including this author) and may require a somewhat refined mathematical treatment, possibly involving notions in Derived Algebraic Geometry. 
This is particularly the case if one wants to map the conformal blocks to objects in the derived categories of D-module used to describe BAA branes in Geometric Langlands or of quasi-coherent sheaves used to describe BBB branes.

A full mathematical treatment of these examples goes beyond the scope of this paper, but we will at least attempt to provide physical motivations for 
 these subtleties. Appendix \ref{app:dmodules} describes in some detail some finite-dimensional examples of D-modules as Chan-Paton bundles for BAA branes in $\C^{2n}$ defined by simple choices of boundary degrees of freedom in the UV. These examples capture some of the subtleties which arise in the construction of conformal blocks. 

Finally, we suspect that the VOAs discussed in this paper can actually occur as algebras of BPS local operators 
on certain deformed supersymmetric boundary conditions for the 3d $\CN=4$ gauge theories. This is the case for 
theories of free hypermultiplets \cite{Costello:2018fnz,Gaiotto:2017euk} and should remain true when gauge fields are added to the mix. 
This would give a direct physical motivation for the relationship between 
the space of conformal blocks of the VOAs and the space of super-symmetric ground states for the corresponding 3d SQFTs.

\section{3d $\CN=4$ gauge theories on a Riemann surface}
Recall that the three-dimensional $\CN=4$ supersymmetry algebra admits an $SU(2)_H \times SU(2)_C$ R-symmetry group.
In standard Lagrangian theories the $SU(2)_H$ R-symmetry group acts on the hyper-multiplet scalar fields
while $SU(2)_C$ acts on the vectormultiplet scalar fields. Three-dimensional mirror symmetry 
exchange the $SU(2)_{H,C}$ subgroups.

In order to define a super-symmetric  compactification on a Riemann surface, we have two natural choices: 
we can twist by the Cartan subgroup of $SU(2)_H$ or by the Cartan subgroup of $SU(2)_C$. 
We denote the two possibilities as ``H-twist'' or ``C-twist'' respectively. When the 3d theories are used 
to define enriched Neumann boundary conditions for four-dimensional SYM, as in \cite{Gaiotto:2016aa}, these twists 
can also be denoted respectively as a ``BAA twist'' and ``BBB twist'',  according to the type of 
branes they give rise to. 

\subsection{The $H$-twist}
Consider a three-dimensional $\CN=4$ SQFT $T$ with an unbroken $SO(2)_H$ Cartan subgroup of the $SO(3)_H$ R-symmetry
which rotates the three complex structures on the Higgs branch. We are mainly interested in renormalizable $\CN=4$ gauge theories, 
which satisfy this requirement automatically in the absence of complex FI parameters. 

The $SO(2)_H$ symmetry can be used to compactify the theory on a Riemann surface $\CC$ while preserving four scalar supercharges.
The result is an effective $\CN=4$ supersymmetric quantum mechanics which we can denote as the $H$-twist of $T$.
We denote as $\CH_H[T,\CC]$ the space of supersymmetric ground states of this quantum mechanics. 

Our objective is to justify and test the following conjectures: 
\begin{itemize}
\item The space $\CH_H[T,\CC]$ can be identified with the space of conformal blocks on $\CC$ for a vertex algebra $\bA_H[T]$. 
\item If $G_H$ is the group of global symmetries acting on the Higgs branch of $T$, $\CH_H[T,\CC]$ can be promoted to a (twisted) D-module over 
the space $\Bun[\CC,G_H]$ of $G_H$ bundles on $C$. 
The D-module structure is associated to the presence of a $G_H$ current subalgebra in $\bA_H[T]$.
\item If $G_C$ is the group of global symmetries acting on the Coulomb branch of $T$, $\CH_H[T,\CC]$ can be promoted to a sheaf over 
the space $\Loc[\CC,G_C]$ of (complexified) $G_C$ flat connections on $C$. 
The sheaf structure is associated to an outer $G_C$ automorphism of $\bA_H[T]$ which allows one to couple it to $G_C$ flat connection. 
\footnote{Here we identified the space of complexified $G_C$ flat connections with the space of local systems $\Loc[\CC,G_C]$. 
The two spaces are topologically the same but not algebraically. The difference is important 
in the mathematical treatment of the Geometric Langlands program. The conformal blocks for $\bA_H[T]$ are ultimately defined 
as solutions of Ward identities which depend polynomially on the complexified $G_C$ flat connection on $C$,
seen as a holomorphic connection on a bundle. 
As far as we understand, that means that the conformal blocks are naturally associated to the space of complexified $G_C$ flat connections 
on $C$ rather than the space of local systems.} 

\end{itemize}

The theory $T[G]$ which appears in the study of $S$-duality in four-dimensional $\CN=4$ $G$ gauge theory has 
$G_H = G$ and $G_C = {}^\vee G$, Geometric Langlands dual groups. We expect the algebra $\bA_H[T[G]]$ to contain a $G$ current algebra and have a ${}^\vee G$ outer automorphism.
Correspondingly, the space of ground states $\CH_H[T[G],\CC]$ should be simultaneously a (twisted) D-module over $\Bun[\CC,G]$ and a sheaf over $\Loc[\CC,{}^\vee G]$.
The relation between S-duality and Geometric Langlands duality suggests that $\CH_H[T[G],\CC]$
should play the role of a ``duality kernel'' in the Geometric Langlands program. 

We propose an explicit construction of the algebra $\bA_H[T]$ when $T$ is a gauge theory with gauge group $G$ and matter hypermultiplets in a 
symplectic representation $M$ of $G$: we define $\bA_H[T]$ as an (extended) coset model given by the quotient of a theory of free symplectic bosons 
valued in $M$ by the $G$ current subalgebra generated by the moment maps in $M$. 

Concretely, the symplectic bosons vertex algebra $\Sb[M]$ is generated by holomorphic fields $Z_a$ of spin $1/2$ valued in $M$, 
with OPE controlled by the symplectic form $\omega_{ab}$ on $M$:
\begin{equation}
Z_a(z) Z_a(w) \sim \frac{\omega_{ab}}{z-w}
\end{equation}
The vertex algebra contains WZW currents valued in the Lie algebra of $G$:
\begin{equation}
J_I(z) = :\mu_G^I(Z): \equiv : Z_a(z) T_I^{ab} Z_b(z):
\end{equation}
where $T_I^{ab}$ generate the symplectic action of $G$ on $M$. 

The vertex algebra $S[M]$ decomposes into a direct sum of representations of the $\hat G$ current algebra generated by the WZW currents $J_I(z)$.
In the absence of Abelian factors in the gauge group, we define the coset vertex algebra $\bA_H[T]$ as the coefficient of the vacuum Verma module in the sum:
\begin{equation} \label{eq:coset}
\Sb[M] = \left[ \bA_H[T] \otimes V_0 \right] \oplus \cdots
\end{equation}
This is what is usually called a coset in the physics literature and denoted as 
\begin{equation}
\bA_H[T] = \frac{\Sb[M]}{\hat G}
\end{equation}
In unitary theories, the coset is usually computed by looking at vertex operators in the original theory with trivial OPE with the WZW currents $J_I(z)$. 
Because the symplectic boson vertex algebra is not unitary, it is possible for the vacuum module to appear a submodule of a larger indecomposable module
and thus our definition of coset turns out to be a bit more restrictive than that. We will see an explicit example later on.

In the presence of Abelian factors in the gauge group, the vertex algebra $\bA_H[T]$ will be graded by characters for the Abelian gauge symmetry. 
The degree $0$ part is defined as before. The degree $p$ part is defined as the coefficient of a Verma module of momentum $p$ for the Abelian currents, vacuum for the non-Abelian currents. 
\begin{equation}
\Sb[M] = \oplus_p \left[ \bA_H^{(p)}[T] \otimes V_p \right] \oplus \cdots
\end{equation}
In order to get a standard (fermionic) VOA, we should restrict the momentum $p$ to values for which the conformal dimensions of the 
coset fields are (half) integral. Typically, this will mean $p$ lies in some full rank sublattice of the charge lattice, as the level of Abelian WZW currents 
in the symplectic boson theory will be integral. 

In any case, we expect the coset algebra to contain WZW currents $:\mu_{G_F}(Z):$ valued in the Lie algebra of the Higgs branch flavor symmetry $G_F$. 
Furthermore, the $p$ grading of the algebra gives the expected action of the Abelian subgroup of $G_C$ which is visible in the UV description of $\CN=4$ gauge theories: each $U(1)$ factor in $G$ contributes a $U(1)$ factor to $G_C$.  

In many cases, the $G_C$ global symmetry of $\CN=4$ gauge theories is enhanced in the IR to a larger non-Abelian symmetry. A crucial check of our conjecture will be the existence of a corresponding enhancement of the global symmetry of $\bA_H[T]$.

In many important situations, where the level of the $\hat G$ current algebra is sufficiently negative and integral, we have found that a certain operation of BRST reduction provides similar results as the 
coset operation and may even be better motivated conceptually. It will allow us to make contact with the work of \cite{Beem:2013aa,Beem:2014aa}
and borrow very useful results about hidden symmetries of certain VOAs. 

\subsection{The $C$-twist}
Mirror symmetry exchanges the role of $SO(3)_H$ and $SO(3)_C$, the R-symmetry which rotates the three complex structures on the Coulomb branch.
A twisted compactification of $T$ on $\CC$ which employs the Cartan subgroup $SO(2)_C$ of $SO(3)_C$ gives an 
effective $\CN=4$ supersymmetric quantum mechanics which we can denote as the $C$-twist of $T$. 
We denote as $\CH_C[T,\CC]$ the space of supersymmetric ground states of this quantum mechanics. 

The following statements should hold true: 
\begin{itemize}
\item The space $\CH_C[T,\CC]$ can be identified with the space of conformal blocks on $\CC$ for a vertex algebra $\bA_C[T]$. 
\item If $G_C$ is the group of global symmetries acting on the Higgs branch of $T$, $\CH_C[T,\CC]$ can be promoted to a (twisted) D-module over 
the space $\Bun[\CC,G_C]$ of $G_C$ bundles on $C$. 
The D-module structure is associated to the presence of a $G_C$ current subalgebra in $\bA_C[T]$.
\item If $G_H$ is the group of global symmetries acting on the Coulomb branch of $T$, $\CH_C[T,\CC]$ can be promoted to a sheaf over the space $\Loc[\CC,G_H]$
of (complexified) $G_H$ connections on $C$. 
The sheaf structure is associated to outer $G_H$ automorphism of $\bA_C[T]$ which allows one to couple it to $G_H$ flat connections. 
\end{itemize}
For theories of free hypermultiplets, we have a simple prescription for $\bA_C[T]$ as an algebra $\Fc[M]$ of {\it fermionic currents}, described by an OPE
\begin{equation}
j_a(z) j_b(w) \sim \frac{\omega_{ab}}{(z-w)^2}
\end{equation}

We do not have a prescription for computing $\bA_C[T]$ for general gauge theories. An obvious strategy, when possible, is to look for a 
mirror gauge theory description $T^!$ of $T$ and compute $\bA_H[T^!]$ instead. 

\subsection{Sheafs and D-modules from $tt^*$ geometry}
Here we would like to briefly explain the reason for the appearance of $\Bun[\CC,G_H]$ and $\Loc[\CC,G_C]$ 
in our story. We refer the reader to \cite{Gaiotto:2016aa} for a more detailed discussion. 

The four supercharges we consider can be better understood by referring to the mirror Rozansky-Witten (mRW) twist of the 3d theory, which is a topological twist 
in three dimensions. In our twisted compactification on $C$, the mRW supercharge can be decomposed into two parts with opposite $SO(2)_H$ charge: 
\begin{equation}
Q_{mRW} = Q_H + \bar Q_H
\end{equation}
It belong to a general family of nilpotent supercharges 
\begin{equation}
Q_H^\zeta = Q_H + \zeta \bar Q_H
\end{equation}
The existence of this family of supercharges in an $\CN=4$ supersymmetric quantum mechanics 
constrains how the space of ground states is fibered over certain parameter spaces of supersymmetric deformations of 
quantum mechanics. These constraints on the Berry connection were first discussed in the study of the $tt^*$ geometry 
of $(2,2)$ two-dimensional sigma models \cite{Cecotti:1991me}.

The specific form of the Berry connection constraints depends on the specific form of the super-multiplet 
to which the deformation of the supercharges and Hamiltonian belongs \cite{Cecotti:2013mba}. The supermultiplet may include several deformations and/or protected operators. 

In general, the Berry connection constraints can be expressed in terms of a ``Lax connection'', a family 
of differential operators $D^\zeta$ on the parameter space which depends holomorphically and linearly in $\zeta$ and commute with each other 
at any given value of $\zeta$. Depending on the specific deformation super-multiplet, the differential operators can take different forms. They are always built from 
the the Berry connections associated to deformations in the supermultiplet and from the expectation values of protected operators in the same 
supermultiplet. 

The original work on $tt^*$ geometry involved supermultiplets which contain a complex deformation parameter and an extra chiral operator, 
such that the differential operators $D^\zeta$ are the Lax connection for a Hitchin system on the parameter space:
\begin{equation}
D^\zeta_u = D_u + \frac{\Phi_u}{\zeta} \qquad \qquad D^\zeta_{\bar u} = D_{\bar u} + \zeta \bar \Phi_{\bar u}
\end{equation}
where $u$ is a complex structure deformation, $D_u$ the associated Berry connection on the space of ground states and $\Phi_u$ the expectation value 
between ground states of the chiral operator associated to the $u$ deformation. We denote these deformation parameters as ``BAA-type'' deformations
as the data can be employed to define BAA branes in 2d $(4,4)$ sigma models. 

%Twisted mass deformations of $(2,2)$ theories give an example of super-multiplets which contain three real deformation parameters and an extra real operator, 
%such that the differential operators $D^\zeta$ are the Lax connection for a 
%system of Bogomolny equations on the parameter space:
%\begin{equation}
%D^\zeta_\C = D_{m} + \frac{D_{\theta} + M}{\zeta} \qquad \qquad D^\zeta_{\bar u} = D_{\bar m} + \zeta \left(-D_{\theta} + M\right)
%\end{equation}
%where $m$ is the twisted mass, $\theta$ is a flavor Wilson line and $M$ is the expectation value of a moment map operator.  

Another important possibility are Cauchy-Riemann equations for tri-holomorphic bundles on hyper-K\"ahler parameter spaces. For example, for an $R^4$ parameter space they would look like 
\begin{equation}
D^\zeta_u = D_u + \frac{D_{\bar v}}{\zeta} \qquad \qquad D^\zeta_{\bar u} = D_{\bar u} - \zeta D_v
\end{equation}
When the theory has several supermultiplets of deformations, the $D^\zeta$ operators all commute with each other at fixed $\zeta$. 
We denote these deformation parameters as ``BBB-type'' deformations as the data can be employed to define BBB branes
in 2d $(4,4)$ sigma models.

There are two natural way to deform our compactification of $T$ on $\CC$, by coupling to background connections for $G_H$ or $G_C$ 
on $\CC$. We can identify supersymmetric deformations by looking at the BPS equations for these background fields. 

There are no constraints on $G_H$ connections, but the dependence on the holomorphic part of the connection is $Q$-exact. 
The topological theory is thus coupled only to a $G_H$ bundle and we get a $\Bun[\CC,G_H]$ factor in the parameter space. 
This type of deformation is analogous to the complex structure deformations in the original $tt^*$ work. The corresponding chiral operator is 
one of the moment map operators on the Higgs branch of the 3d theory. 
The BAA-type structure associated to $G_H$ connections is thus a connection on the sheaf of ground states on $\Bun[\CC,G_H]$
together with a Higgs field, the expectation value of the moment map operator, which thogether satisfy the higher-dimensional version of Hitchin equations on $\Bun[\CC,G_H]$. 

For generic $\zeta$, and in particular for $\zeta=1$, the Lax connection for such Hitchin system is a flat connection on $\Bun[\CC,G_H]$. It equips 
the cohomology of $Q_H^\zeta$ with the structure of a D-module on $\Bun[\CC,G_H]$.
More precisely, we expect that one should be able to identify the output of the supersymmetric quantum mechanics with an object in some appropriate derived category of D-modules on $\Bun[\CC,G_H]$.

The second possibility is a bit more intricate. The BPS equations require us to turn on both a $G_C$ connection $A_C$ and a background complex adjoint scalar $\Phi_C$ 
in the $G_C$ twisted vectormultiplet (i.e. a complex FI parameter).
The twist by $SO(2)_H$ makes the complex FI parameters $\Phi_C$ of $T$ into 
one forms on $\CC$, valued in the Lie algebra of $G_C$. A pair $(A_C, \Phi_C)$ of background $G_C$ connection $A_C$ 
and scalar $\Phi_C$ preserves $Q_H^\zeta$ if the auxiliary Lax connection 
\begin{equation}
D_z^\zeta = D_z[A_C] + \frac{\Phi_{C,z}}{\zeta}  \qquad \qquad D_{\bar z}^\zeta = D_{\bar z}[A_C] + \zeta \Phi_{C,\bar z}
\end{equation}
is flat for all $\zeta$. It preserves all four scalar supercharges if $(A_C, \Phi_C)$ is a solution of $G_C$ Hitchin's equations on $C$. 

Thus this factor of the parameter space is the Hitchin moduli space $\CM[G_C, \CC]$. The corresponding BBB-type structure is a tri-holomorphic sheaf 
on $\CM[G_C, \CC]$: the sheaf of ground states for the $\CN=4$ quantum mechanics has a Berry connection which is holomorphic in all complex structures of $\CM[G_C, \CC]$.
The choice of $\zeta$ is a choice of complex structure on $\CM[G_C, \CC]$.

For generic $\zeta$, and in particular for $\zeta=1$, we can identify the parameter space with $\Loc[\CC,G_C]$. For any given $\zeta$, it gives the cohomology of $Q_H^\zeta$ the structure of an holomorphic sheaf on $\CM[G_C, \CC]$ in complex structure $\zeta$. More precisely, we expect that one should be able to identify the output of the supersymmetric quantum mechanics with an object in some appropriate derived category of sheaves on $\Loc[\CC,G_C]$.

As the Lax connections on the two factors of the parameter space commute, the sheaf structure on $\Loc[\CC,G_C]$ and the D-module structure on $\Bun[\CC,G_H]$
are compatible, i.e. the flat connection on $\Bun[\CC,G_H]$ with parameter $\zeta$ commutes with the anti-holomorphic derivatives in complex structure $\zeta$ for the sheaf on $\Loc[\CC,G_C]$. 

The BAA and BBB structures on the supersymmetric ground states can be encoded as BBB and BAA branes on $\CM[G_C, \CC]$ and $\CM[G_H, \CC]$ respectively. 
Physically, this arises from the promotion of a three-dimensional $\CN=4$ SQFT to a half-BPS interface for four-dimensional $G_H$ and $G_C$ 
$\CN=4$ gauge theories: compactification on a Riemann surface $\CC$ reduces the four-dimensional gauge theories to 
$\CM[G_C, \CC]$ and $\CM[G_H, \CC]$ sigma-models and the 3d interface to a BPS interface between the two sigma models,
which is of BBB type on one side and BAA on the other side. 

In particular, the sheaf of ground states for the $T[G]$ theory gives a BPS interface between the  $\CM[G, \CC]$ and $\CM[G^\vee, \CC]$
sigma models which should implement the mirror symmetry relation between the two sigma models, i.e. the Geometric Langlands duality.

We refer to \cite{Gaiotto:2016aa} for more details and for a description of the geometric structures which emerge at $\zeta=0$. 

\subsection{From hypermultiplets to symplectic bosons and fermionic currents}
In the absence of gauge fields, there is a simple way to understand the algebras $\bA_H[M]$ and $\bA_C[M]$ we associate to 
hypermultiplets valued in $M$. 

\subsubsection{H-twist}
As standard hypermultiplet scalars transform in a doublet of $SO(3)_H$, 
the H-twist makes them into spinors on $\CC$. The hypermultiplet fermions are already spinors on $\CC$ to start with.

The three-dimensional action can be recast as a supersymmetric quantum mechanics akin to a Landau-Ginzburg theory 
with a K\"ahler target manifold \cite{Bullimore:2016aa}. The target of the quantum mechanics is the space of sections $Z$ of the bundle $K^{1/2} \otimes M$ 
on $\CC$. The superpotential is the symplectic boson action:
\begin{equation}
W = \int_{\CC} \langle Z, D_{\bar z} Z \rangle
\end{equation} 
where $\langle \cdot ,\cdot  \rangle$ is the symplectic pairing on $M$ and $D_{\bar z}$ the anti-holomorphic covariant derivative associated to the bundle.  

Notice that if we pick a global symmetry group $G_H$ acting simplectically on the hypermultiplets, 
we can take $M$ to be a non-trivial $G_H$ bundle rather than the constant bundle. Thus $\Bun[\CC,G_H]$
is a parameter space of complex structure/superpotential deformations for the LG quantum mechanics. 
The variation of $W$ along $\Bun[\CC,G_H]$, which is the integral over $\CC$ of the moment map $\mu(Z)$ contracted with the variation of 
the anti-holomorphic connection, 
\begin{equation}
\delta W = \int_{\CC} \mu(Z) \cdot \delta A^{G_H}_{\bar z}
\end{equation} 
gives a local operator in the quantum mechanics which combines with the Berry connection to give the $tt^*$ structure mentioned above.  

The space of ground states of an $\CN=4$ Landau-Ginzburg quantum mechanics with finite-dimensional target space $U$ is the cohomology of $U$ 
relative to the locus where $\mathrm{Re} \, W \ll 0$, i.e. the space of integration cycles for forms which behave as $e^W$. This cohomology has an integral basis and 
it is a locally constant sheaf on the space of complex structure/superpotential deformations for the quantum mechanics: the parallel transport is defined by continuous deformations 
of the integration contours. 

This structure can be recast as a D-module: the D-module associated to the Picard-Fuchs equations satisfied by integrals of the form 
\begin{equation}
\oint_\gamma \omega e^W
\end{equation}
where $\omega$ lies in an appropriate $dW$-deformation of De Rham cohomology. See Appendix \ref{app:dmodules} for several examples. 

The finite-dimensional model suggests that the space $\CH_H[M,\CC]$ should coincide with the space of conformal blocks for a
theory of chiral symplectic bosons, defined by the path integral 
\begin{equation} \label{eq:path}
\int DZ e^{\int_{\CC} \langle Z, D_{\bar z} Z \rangle}
\end{equation}
with a (twisted) D-module action given by the WZW current subalgebra defined by the moment maps
\begin{equation}
J_{G_H} = : \mu(Z) :
\end{equation}

This path integral gives a free vertex algebra $\Sb[M]$ with OPE
\begin{equation}
Z_a(z) Z_b(w) \sim \frac{\omega_{ab}}{z-w}
\end{equation}
where $\omega$ is the symplectic form on $M$. A simple way to understand why symplectic bosons can be coupled to 
a gauge bundle is to observe that this OPE is invariant under holomorphic gauge transformations of the $Z_a(z)$.

There is an alternative perspective which supports this proposal: the theory of free hypermultiplets admits a boundary condition 
which preserves a $(0,4)$ two-dimensional subalgebra of the supersymmetry algebra. A mRW twist of the theory is known to 
lead to a theory of holomorphic symplectic bosons on the boundary \cite{Costello:2018fnz, Gaiotto:2017euk}. This construction thus gives a map from the space of 
conformal blocks for symplectic bosons to the space of states of mRW-twisted free hypermultiplets. 

The conformal blocks for symplectic bosons on a Riemann surface $\CC$ in the absence of 
a background gauge bundle depend on a choice of spin structure $K^{1/2}$. In the presence of a background gauge bundle $E$, 
$Z_a$ transforms as a section of the associated bundle $E_M \otimes K^{1/2}$. \footnote{Rather than considering this as a choice of 
a bundle $E$ and spin structure $K^{1/2}$ it is more natural to take $E_M \otimes K^{1/2}$ to be some sort of generalized Spin$_\C$ 
structure and the space of conformal blocks as a D-module over the moduli space of such structures.}

As long as $E_M \otimes K^{1/2}$ has no global sections, so that the symplectic boson has no zeromodes on $\CC$, the path integral \ref{eq:path}
has an obvious meaning and gives a single conformal block, i.e. a unique solution of the Ward identities for correlation functions of the $Z_a$.
The partition function is the inverse of the square root of the determinant of the $\bar \partial$ 
operator on $E_M \otimes K^{1/2}$. As one approaches the locus where $E_M \otimes K^{1/2}$ has global sections, the partition function will diverge. 

In a component of the space of bundles where the symplectic bosons have generically no zeromodes, 
a naive description of the space of conformal blocks is a rank 1 D-module with a regular singularity 
at the locus where $E_M \otimes K^{1/2}$ has global sections. The finite-dimensional examples in Appendix \ref{app:dmodules} 
make it clear that this description is incomplete and additional conformal blocks are hidden at special loci in 
the space of bundles. Such hidden conformal blocks are even more important in components of the space of bundles 
where zeromodes exist generically. 

These additional conformal blocks are important in matching and improving the classical description \cite{Gaiotto:2016aa} 
of the BAA brane as a complex Lagrangian submanifold of the space of Higgs bundles
$(E,\varphi)$: the Lagrangian has a component wrapping the $\varphi=0$ locus and extra components which sit on the co-normal bundle to the 
locus where $E_M \otimes K^{1/2}$ has global sections.

A general description of the space of conformal blocks is that of a complex of D-modules, with a differential which 
imposes the Ward identities on correlation functions. This is described in Appendix \ref{app:dmodules}. 
We expect that this complex can be systematically simplified, at least locally on the space of bundles, 
but we leave that to future work and focus on concrete examples. 

\subsubsection{C-twist}

The RW twist of a theory of free hypermultiplets (i.e. the mRW twist of a theory of free twisted hypermultiplets)
leaves the hypermultiplet scalars unaffected, but changes the quantum numbers of the fermions: 
part of the fermions become spin-zero superpartners of the bosonic scalar fields and the other half become one-forms. 
The spin zero fields are rather boring, but the fermionic one forms have an interesting Chern-Simons action built from the symplectic pairing on $M$.
It is natural to expect that the space of ground states on $\CC$ will be the space of conformal blocks of fermionic 
WZW currents $\Fc[M]$ valued in $M$, with OPE
\begin{equation} \label{eq:fermionic}
j_a(z) j_b(w) \sim \frac{\omega_{ab}}{(z-w)^2}
\end{equation}

Notice that if we pick a global symmetry group $G_C$ acting symplectically on the hypermultiplets, this system has no $G_C$-valued WZW subalgebra: 
we can couple the system to a flat $G_C$ connection, but 
there is no holomorphic current to encode the infinitesimal changes in the connection. The conformal blocks form a sheaf over
$\Loc[\CC,G_C]$, as expected. 

A simple way to understand this fact is to observe that the OPE \ref{eq:fermionic}
is not invariant under holomorphic gauge transformations, because of the double pole. It can be made invariant by adding a dependence on an 
holomorphic $G_C$ connection ${\cal A}_{ab}$ on $C$
\begin{equation} \label{eq:fermionic2}
j_a(z) j_b(w) \sim \frac{\omega_{ab}}{(z-w)^2} + \frac{{\cal A}_{ab}(w)}{z-w}
\end{equation}
which combined with the bundle data into a holomorphic description of a $G_C$ local system, i.e. a bundle equipped with a holomorphic connection on the Riemann surface $C$.  

The spin zero fields in the hypermultiplets are expected to be completely trivial as long as the $G_C$ local system has no scalar global sections \cite{Gaiotto:2016aa}. 
If scalar global sections exist, the system becomes more complicated, in a manner we now describe. Notice that if the 3d theory is coupled to four-dimensional gauge theory,
the vevs for the spin zero fields in the hypermultiplets trigger vevs for the four-dimensional scalar fields which are not included in the 
picture of a sigma model on the Hitchin moduli space. 

The calculations in \cite{Gaiotto:2016aa} predict that the sheaf of supersymmetric ground states should arise from the 
quantization of a phase space given by the de Rahm cohomology of forms on $C$ valued in $M$. 
This is known to coincide with the sheaf of (derived) conformal blocks for 
fermionic currents valued in $M$ \cite{kevinsb}. 

In order to understand the relationship, we can pick a polarization in the phase space which splits into 
$(*,0)$ forms and $(*,1)$ forms and build a Fock space out of $\Omega_{*,0}$.
The Fock space can be identified with a collection of potential correlation functions 
\begin{equation}
\langle j_{a_1}(z_1) \cdots j_{a_n}(z_n) \phi_{b_1}(w_1) \cdots \phi_{b_m}(w_m)\rangle 
\end{equation} 
and the BRST differential takes the schematic form 
\begin{align}
Q \langle \cdots \rangle &=  \int dz d \bar z \langle \left(\bar \partial j_a(z) -  \omega_{ab} \partial \frac{\delta}{\delta j_b(z)} \right) \frac{\delta}{\delta \phi_a(z)} \cdots \rangle + \cr
&+ \int dz d \bar z \langle \left(j_a(z) \omega^{ab} \bar \partial \phi_b(z)+ \partial \phi_b(z) \frac{\delta}{\delta j_b(z)} \right) \cdots \rangle 
\end{align} 

This seems a reasonable definition for a space of (derived) conformal blocks for the fermionic currents. 
The $Q$ cohomology in cohomological degree $0$ consists of correlation functions for the $j_a$ currents satisfying the Ward identities of fermionic currents
\begin{equation}
\langle \left(\bar \partial j_a(z) -  \omega_{ab} \partial \frac{\delta}{\delta j_b(z)} \right) \cdots \rangle =0
\end{equation}
If there are no zeromodes for the scalars, we expect this to exhaust the cohomology. 

\subsection{Coset versus BRST reduction}
The vertex algebra of symplectic bosons is a crucial ingredient of another construction which associates vertex algebras to 
gauge theories with eight supercharges: the construction of vertex algebras for $\CN=2$ four-dimensional SCFTs \cite{Beem:2013aa,Beem:2014aa}. The requirement of conformal symmetry in four dimensions
imposes strong constraints on the gauge theory matter content: the level of the $\hat G$ current algebra in $\Sb[M]$ should be twice the critical level. 

At this particular value of the level, it is possible to pair up the $S[M]$ algebra with a system of $b,c$ ghosts valued in the gauge Lie algebra
and write down a BRST operator of the schematic form $Q_{BRST} = c J + bcc$. The BRST cohomology produces the 
vertex algebras for the $\CN=2$ four-dimensional gauge theory. 

It is quite obvious that if we build a 3d theory $T$ with the same matter content and gauge group as a 4d SCFT, operators in our coset 
will belong also to the $Q_{BRST}$ cohomology. Moreover, the central charge of the resulting VOAs also coincide: 
the central charge for the ghosts precisely cancels the central charge of the WZW currents 
at twice the critical level:
\begin{equation}
\frac{k \dim  G}{k+h}\bigg|_{k = - 2 h} + (-2) \times \dim  G= 0
\end{equation}
Inspection of examples will strongly suggest that $\bA_H[T]$ coincides in this situation with the 4d chiral algebra. We conjecture 
\begin{equation}
\bA_H[T] = \frac{\Sb[M]}{\hat G_{- 2 h}} = \left\{ \Sb[M] \times (b,c), Q_{BRST} \right\}
\end{equation}

Assuming that this correspondence holds will be rather useful later in the paper: the 4d chiral algebra of theories of class $S$ has unexpected symmetries 
which are thus inherited by our coset and which will be instrumental in demonstrating the Coulomb branch symmetry enhancements for unitary quiver gauge theories.

If the matter content of the three-dimensional theory is beyond the amount allowed in four-dimensions, 
so that the level of the WZW currents is more negative than twice the critical level, by an integral amount $-n$,
we can still {\it add} a standard $\hat G_n$ WZW model to the symplectic bosons and then apply the BRST reduction. 

Again, the resulting VOA seems closely related to the one obtained by a direct coset of the symplectic bosons. 
For example, the central charge of the $\hat G_n$ WZW model combines with the central charge of the ghosts 
to cancel the central charge of the $\hat G$ WZW currents in the symplectic boson theory: 
\begin{equation}
\frac{k \dim  G}{k+h}\bigg|_{k = - 2 h-n} + \frac{k \dim  G}{k+h}|_{k = n} + (-2) \times \dim  G = 0
\end{equation}
Again, we expect this BRST construction to give an alternative definition of  $\bA_H[T]$. We conjecture 
\begin{equation}
\bA_H[T] = \frac{\Sb[M]}{\hat G_{- 2 h-n}} = \left\{ \Sb[M] \times \hat G_n \times(b,c), Q_{BRST} \right\}
\end{equation}

\section{Free hypermultiplets}
We will discuss now some examples of VOA associated to free hypermultiplets in various representations. 
\subsection{H-twist of a single hypermultiplet}
The vertex algebra $\Sb[\C^2]$ of a single symplectic boson has two bosonic generators, $X(z)$ and $Y(z)$, with OPE
\begin{equation}
X(z) Y(w) \sim \frac{1}{z-w}
\end{equation}
and conformal dimension $1/2$. Several of the features we discuss below can be found discussed at length in \cite{Lesage:2002ch}.

The stress tensor can be written as 
\begin{equation}
T = \frac{1}{2} X \partial Y - \frac{1}{2} Y \partial X
\end{equation}
and gives a central charge of $c_{XY}=-1$. \footnote{We can check that this is the correct stress tensor
\begin{equation}
T(z) X(w) \sim \frac{1}{2} \frac{X(w)}{(z-w)^2}  + \frac{\partial X(w)}{z-w} \qquad \qquad T(z) Y(w) \sim \frac{1}{2} \frac{Y(w)}{(z-w)^2}  + \frac{\partial Y(w)}{z-w}
\end{equation}
and compute the central charge 
\begin{equation}
T(z) T(w) \sim -\frac{1}{2}\frac{1}{(z-w)^4} + \frac{2 T(w)}{(z-w)^2} + \frac{\partial T(w)}{z-w}
\end{equation}
}

The basic vacuum module of the symplectic boson VOA is generated by half-integral modes in the expansion 
\begin{equation}
X(z) = \sum_{n=-\infty}^\infty \frac{X_{n-\frac12}}{z^n} \qquad \qquad Y(z) = \sum_{n=-\infty}^\infty \frac{Y_{n-\frac12}}{z^n} 
\end{equation}
with $[X_{n-\frac12}, Y_{m-\frac12}]=\delta_{n+m,1}$. The module is generated from the identity by the action of the negative modes in the expansion.

The vacuum module belongs to the sector with Neveu-Schwarz boundary conditions for $X$ and $Y$. The sector with Ramond boundary conditions 
is somewhat more subtle, because of the existence of zeromodes 
which satisfy an Heisenberg algebra: 
\begin{equation}
X(z) = \sum_{n=-\infty}^\infty \frac{X_{n}}{z^{n+\frac12}} \qquad \qquad Y(z) = \sum_{n=-\infty}^\infty \frac{Y_{n}}{z^{n+\frac12}} 
\end{equation}
with $[X_{n}, Y_{m}]=\delta_{n+m,0}$ and in particular $[X_0,Y_0] = 1$. Useful Ramond modules can be induced 
from any modules for the Heisenberg algebra of zeromodes.

Obvious choices are modules generated from vectors $|R,\pm\rangle$ 
which are annihilated either by $X_0$ or $Y_0$ and all positive modes. 
A less obvious choice is a module generated by vectors $|R,\lambda+n\rangle$ annihilated by positive modes, 
with $n$ an integer and $0 < \lambda<1$ and  
\begin{equation}
Y_0 |R,\lambda+n\rangle =  |R,\lambda+n+1\rangle \qquad \qquad X_0 |R,\lambda+n\rangle =  (n + \lambda) |R,\lambda+n-1\rangle
\end{equation} 
As our focus in this paper is on Riemann surfaces with no punctures, vertex operators for general modules of the symplectic boson algebra 
will play a limited role. 

As the symplectic boson CFT can be described by a free chiral action $\int_C X \bar \partial Y$, 
we expect that as long as the action has no zeromodes the space of conformal blocks will be one-dimensional, generated 
by the Gaussian path integral with that action. In particular, the partition function should be just the inverse of the determinant of the $\bar \partial$ operator
acting on sections of $K^\frac12$. If we do not couple the symplectic boson to a background gauge bundle, we need to select an even spin structure 
$K^\frac12$ in order to avoid zeromodes. It is more natural, though, to couple the system to background gauge fields. We will come back to that momentarily.

This expectation can be verified by directly solving on the Riemann surface $C$ the Ward identities of the symplectic boson VOA 
or by assembling the conformal block by sewing up punctured spheres. The Ward identities express the correlation functions of $X$ and $Y$ fields
in terms of the Green's function for the $\bar \partial$ operator. Concretely, $\frac{ \langle X(z) Y(w) \rangle_{C} }{\langle 1 \rangle_{C}}$ is the unique meromorphic section of $K^{\frac12}$ 
with a single pole of residue $1$ at $w$. The overall normalization is determined by computing from the Green's function the stress tensor one-point function $\frac{ \langle T(z) \rangle_{C} }{\langle 1 \rangle_{C}}$
and thus the dependence on the complex structure of $C$. 

Similarly, the Ward identities allow one to reduce any sphere three-point function 
of vacuum descendants to the sphere partition function. Conformal blocks on a generic Riemann surface equipped with an even spin structure can be computed, say,
by sewing together pairs of punctures from a sphere with $2g$ NS punctures by inserting complete sets of descendants of the identity,
possibly with a twist acting as $-1$ on $X$ and $Y$ in order to select a specific spin structure. It is also possible to 
reproduce the answers by sewing along Ramond sector channels, but there are important subtleties associated to the 
zeromodes.  

If we consider odd spin structures, instead, we cannot define a partition function unless we remove the 
zeromodes. It is possible to remove the zeromodes without locally interfering with the Ward identities,  
but the global behaviour of correlation functions is spoiled by logarithmic monodromies. The situation is improved by 
introducing a $U(1)$ gauge bundle. We will do that now.  

\subsection{Current subalgebras: $U(1)_{-1}$}
The symplectic boson theory has an obvious $U(1)_{-1}$ WZW current 
\begin{equation}
J = X Y
\end{equation}
rotating $X$ and $Y$ with charge $\pm1$. This allows one to couple the symplectic boson to 
a $U(1)$ bundle $L$ on the Riemann surface. 

Concretely, this statement is related to the observation that the OPE of symplectic bosons is well-behaved under 
holomorphic gauge transformations: as the OPE has a single pole, the replacement $X(z) \to g(z) X(z)$
and $Y(z) \to g(z)^{-1} Y(z)$ does not change the singular part of the OPE, and shifts $J$ by the expected $g^{-1} \partial g$ determined by the anomaly. 

Notice that it is natural to think about the $L \otimes K^{\frac12}$ bundle as a Spin$_\C$ structure on the Riemann surface, 
rather than choosing a spin structure and then a line bundle. Correspondingly, the conformal blocks are best defined 
over the moduli space of Spin$_\C$ structures on the Riemann surface. 

If we take the line bundle to have degree $0$, the path integral produces a partition function 
\begin{equation}
Z_{C,L} = \frac{1}{\det \bar \partial_{L \otimes K^{\frac12}}}
\end{equation}
which has a pole along the $\Theta$-divisor in the space of line bundles, where zeromodes appear. 

In the first approximation, we can envision the conformal blocks in degree $0$ as a one-dimensional 
line bundle on the space of $U(1)$ bundles $\Bun_0(U(1),C)$ equipped with the structure of a (twisted) D-module 
with a regular singularity at the $\Theta$-divisor. The partition function 
plays the role of a flat section of that D-module.
Recall that the D-module structure on conformal blocks is simply the statement that we can change the $U(1)$ bundle infinitesimally by inserting a $U(1)$ current  
in the partition function, decomposing it into the OPE of two symplectic bosons and use the Ward identities to re-express that in terms of the original partition function. 

The discussion in Appendix \ref{app:dmodules} makes it clear that such a description, though, is dangerously simplistic. 
The space of conformal blocks should really be thought of as a complex of (infinite-dimensional) vector bundles with a D-module structure.
The cohomology of that complex away from the $\Theta$-divisor is the naive one-dimensional space of conformal blocks,
but a lot of structure and hidden components may be present at the $\Theta$-divisor itself. 

There is a simple trick which produces examples of such non-trivial components of the space of conformal blocks:
start from the standard partition function and correlation functions and take a discontinuity across the $\Theta$-divisor,
transforming the poles into delta functions. Equivalently, we can act
on the standard conformal blocks with a $\partial^\dagger$ operator along $\Bun_0(U(1),C)$,
which again transforms poles into delta functions. This agrees with the expectation from finite-dimensional analogue systems in Appendix \ref{app:dmodules}
that extra non-trivial components may be found in cohomological degree $-1$ at loci where a pair of dual zeromodes appear. 

The example of genus $1$ conformal blocks is already rather instructive. For a generic point in $\Bun_0(U(1),E_\tau)$
parameterized by the variable $x$, the partition function and correlation functions take the form \footnote{To check this formulae, observe that $\langle X(z) Y(w) \rangle$ is the unique meromorphic section with a single pole 
or residue $\langle 1 \rangle$ and that it gives the correct stress-tensor 1-pt function proportional to $\partial_\tau \langle 1 \rangle$. }
\begin{align}
\langle 1 \rangle &= \frac{\eta(\tau)}{\theta(x,\tau)} \cr
\langle X(z) Y(w) \rangle &=2 \pi i  \frac{\eta(\tau)^4}{\theta(x,\tau)^2}\frac{\theta(z-w+x,\tau)}{\theta(z-w,\tau)} \cr
\cdots
\end{align}
Taking the discontinuity at $x=0$ we get our candidate hidden conformal block:  
\begin{align}
\langle 1 \rangle &= \frac{1}{\eta^2(\tau)} \delta(x) \cr
\langle X(z) Y(w) \rangle &=\frac{1}{\eta^2(\tau)} \delta'(x) + \frac{1}{\eta^2(\tau)}\frac{\theta'(z-w,\tau)}{\theta(z-w,\tau)} \delta(x) \cr
\cdots
\end{align}
The ``partition function'' is a natural regularization of the naive path integral, with zero-modes removed. The correlation function 
may appear worrisome because $\frac{\theta'(z-w,\tau)}{\theta(z-w,\tau)}$ shifts by a constant as $z \to z - \tau$. This compensates, though, 
the fact that $e^{2 \pi i x} \delta'(x) = \delta'(x) - 2 \pi i \delta(x)$. Thus the correlation function is still a section of the correct bundle. 

Similar considerations apply in higher genus, though new components in even lower cohomological degree may appear 
at special loci in the $\Theta$ divisor. 

If the $U(1)$ line bundle has degree greater than $0$, $Y(z)$ will generically have $d$ zeromodes while the equations of motion 
for $X(z)$ will be obstructed. The opposite occurs in negative degree. Solving the Ward identities will simply be generically impossible and the (cohomology of the complex of) 
conformal blocks will be generically trivial. The D-module of conformal blocks, though is still non-trivial: the finite-dimensional example in Appendix \ref{app:dmodules} suggests that non-trivial solutions of Ward identities 
appear at the co-dimension $d+1$ locus $\Theta_d$ where $X(z)$ acquires at least one zeromode and thus $Y(z)$ has $d+1$ zeromodes. 

Concretely, the correlation functions are expected to vanish unless we have $d$ more $X$ insertions than $Y$ insertions. 
At the co-dimension $d+1$ locus $\Theta_d$ in $\Bun_d(U(1),E_\tau)$ where $X(z)$ has some zeromode $\rho(z)$, we can postulate 
\begin{equation}
\langle X(z_1) \cdots X(z_d) \rangle = \rho(z_1) \cdots \rho(z_d) \delta^{(d)}_{\Theta_d}
\end{equation}
Because of the existence of $d+1$ obstructions for the equations of motion of $X$, we can only find the Green's function if we 
allow for logarithmic monodromies in $d$ directions, as on the degree $0$ case. We expect to be able to compensate for that 
using the $d$ normal derivatives of the $\delta_{\Theta_d}$, as before. Thus the next non-trivial correlation function will have a schematic form
\begin{equation}
\langle X(z_1) \cdots X(z_d) X(z_{d+1}) Y(w) \rangle = \sum_a \prod_{b \neq a} \rho(z_b) \left( G(z_a, w) \delta^{(d)}_{\Theta_d}+ g(z_a,w) \cdot \partial \delta^{(d)}_{\Theta_d} \right) 
\end{equation}
etcetera. 

Notice that both the $\Theta$-divisor and the $\Theta_d$ loci for $g-1>d>0$ can be parameterized nicely by the divisor given by the $g-1-d$ zeroes of the $X(z)$ zeromode.
At degree $d = g-1$ the locus $\Theta_{g-1}$ consists of the trivial bundle only and the $X$ zeromode is constant. 
For $d$ greater than $g-1$ we do not expect any interesting conformal blocks. This agrees with the classical picture described in \cite{Gaiotto:2016aa}. Similar considerations apply for negative $d$. 

\subsubsection{Twisted modules}
In the presence of a background $U(1)$ connection one can consider twisted sectors 
for the symplectic boson, where the mode expansion is shifted appropriately
\begin{equation}
X(z) = \sum_{n=-\infty}^\infty \frac{X_{n+\alpha-\frac12}}{z^{n+\alpha}} \qquad \qquad Y(z) = \sum_{n=-\infty}^\infty \frac{Y_{n-\alpha-\frac12}}{z^{n-\alpha}} 
\end{equation}
For $\alpha \neq \frac12$ we have natural highest weight modules annihilated by the positive modes. 
These modules will appear, say, when we sew up a Riemann surface in a gauge where $X$ and $Y$ have non-trivial periodicity 
around the handles. For future reference, we can compute 
\begin{equation}
\langle \alpha| X(z) Y(w) |\alpha \rangle = \frac{w^\alpha}{z^\alpha} \frac{1}{z-w}
\end{equation}
leading to the $U(1)$ charge and scaling dimension of the highest weight vectors
\begin{equation}
\langle \alpha| J(z) |\alpha \rangle = -\frac{\alpha}{z} \qquad \qquad \langle \alpha| T(z) |\alpha \rangle = -\frac{\alpha^2}{2z^2}
\end{equation}
These twisted modules can be obtained from the standard vacuum module by a singular gauge transformation. 

At $\alpha = \frac12$ we need to consider the various possible Ramond sector modules. Approaching $\alpha = \frac12$
from above or below one gets the $|R,\pm\rangle$ modules. Instead the general Ramond modules give us 
\begin{equation}
\langle R,\lambda | X(z) Y(w) |R,\lambda\rangle = \frac{w^{\frac12}}{z^{\frac12}} \frac{1}{z-w} + \frac{\lambda}{z^{\frac12}w^{\frac12}}
\end{equation}
and thus 
\begin{equation}
\langle R,\lambda| J(z) |R,\lambda\rangle = (\lambda -\frac12) \frac{1}{z} \qquad \qquad \langle R,\lambda| T(z) |R,\lambda\rangle = -\frac{1}{8 z^2}
\end{equation}
This module is not obtained by a singular gauge transformation of the vacuum module. We expect it to play an important role in the sewing 
construction of the non-standard conformal blocks described above. It also plays an important role in 
the bosonization of the $XY$ system, which will be a crucial ingredient in the study of 
Abelian mirror symmetry and S-duality.

For example, the characters and traces over the modules generated from $|\alpha \rangle$ or $|R,\pm\rangle$ all essentially give the same $\frac{\eta(\tau)}{\theta(x,\tau)}$
torus partition function and associated correlation functions, with $x = \tau (\alpha-\frac12)+\beta$ and $\beta$ being the $U(1)$ fugacity. 
On the other hand, the characters and traces of the modules 
generated from $|R,\lambda\rangle$ give the $\frac{1}{\eta^2(\tau)} \delta(x)$ torus partition function and associated correlation functions. 

\subsubsection{Current subalgebras: $SU(2)_{-\frac{1}{2}}$}
The symplectic boson vertex algebra actually contains a full set of $SU(2)_{-\frac{1}{2}}$ WZW currents: \footnote{
We can verify the level from the OPEs:
\begin{align}
J^3(z) J^3(w) &\sim -\frac{1}{4} \frac{1}{(z-w)^2} \cr
J^3(z) J^\pm(w) &\sim \pm \frac{J^\pm}{z-w} \cr
 J^-(z) J^+(w) &\sim \frac{1}{2} \frac{1}{(z-w)^2} + \frac{2J^3}{(z-w)}
\end{align}
}
\begin{equation}
J^- = \frac{1}{2} X^2 \qquad \qquad J^3 =\frac{1}{2} X Y \qquad \qquad J^+ = \frac{1}{2} Y^2
\end{equation}

Notice that $X$ and $Y$ can be identified with the spin $\frac{1}{2}$ primaries $Z_\alpha$ for the $SU(2)$ current algebra:
the dimension of a spin $\frac{1}{2}$ primary is precisely $\frac{1}{2}$. Furthermore, the Sugawara stress tensor can be computed from 
\begin{equation}
: J^3 J^3 : = \frac{1}{4} X^2 Y^2 +\frac{1}{4} X \partial Y - \frac{1}{4} Y \partial X
\end{equation}
and 
\begin{equation}
: J^- J^+ : = \frac{1}{4} X^2 Y^2 +Y \partial X \qquad : J^+ J^- : = \frac{1}{4} X^2 Y^2 - X \partial Y
\end{equation}
so that 
\begin{equation}
: J^3 J^3 : -\frac{1}{2}: J^- J^+ : - \frac{1}{2}: J^+ J^- := \frac{3}{2} T
\end{equation}
Thus $T$ coincides with the Sugawara stress tensor for $SU(2)_{-\frac{1}{2}}$. 

The currents of integral spin in the symplectic boson current algebra can be organized into the
vacuum module of $SU(2)_{-\frac{1}{2}}$, while the currents of half-integral spin can be organized into the spin $\frac12$ module of 
$SU(2)_{-\frac{1}{2}}$. Thus we can envision the symplectic boson VOA as an extension of the $SU(2)_{-\frac{1}{2}}$
WZW VOA. 

We can use the $SU(2)_{-\frac{1}{2}}$ WZW symmetry to couple the symplectic boson system to 
$SU(2)$ bundles. Again, it is actually most natural to couple the symplectic boson to an $SU(2)$ version of a Spin$_{\C}$ structure:
rather than picking an $SU(2)$ bundle $E$ and combining it with a spin structure, we can give the product 
$E \otimes K^{\frac12}$ an intrinsic meaning.  This should correspond to $E$ being a section of a certain gerbe. 

From the four-dimensional perspective, this is due to the $Z_2$ anomaly of a single half-hypermultiplet coupled to $SU(2)$ 
gauge fields, which is cancelled by anomaly inflow from a non-trivial discrete theta angle in the four-dimensional bulk. 
The bulk theory with such a discrete theta angle is conventionally denoted as $Sp(1)'$ and is mapped to itself by S-duality. 
Correspondingly, the space of solutions of Hitchin equations twisted by that gerbe should be self-mirror. 

The partition function of the symplectic boson coupled to the twisted $SU(2)$ bundle 
is 
\begin{equation}
Z_{C,E} = \frac{1}{\sqrt{\det{ \bar \partial_{E \otimes K^{\frac12}}}}}
\end{equation}
and has square-root singularities at the co-dimension 1 locus $\Theta_{SU(2)}$ where a zeromode appears. 
Notice that there is a $Z_2$ symmetry mapping $Z_\alpha(z) \to - Z_\alpha(z)$ 
and solutions of Ward identities built from this partition function involve correlation functions with an even number of $Z_\alpha$ insertions.

At that locus $\Theta_{SU(2)}$ we expect to also find a second conformal block (in cohomological degree $0$, see examples in Appendix \ref{app:dmodules}) which has zero partition function, but 
non-zero 1-point function 
\begin{equation}
\langle Z_\alpha(z) \rangle = \rho_\alpha(z) \delta_{\Theta_{SU(2)}}
\end{equation}
proportional to the zeromode $\rho_\alpha(z)$ and more general correlation functions of an odd number of fields involving the $\delta$ function
at $\Theta_{SU(2)}$ and its derivatives. 

\subsection{Free hypermultiplets in a fundamental representation, H-twist}
As a preparation for later sections, we should discuss briefly some features of the VOA $\Sb[\C^{2N}]$ obtained as the product of $N$ copies 
symplectic boson VOAs. 

The vertex algebra has $2N$ bosonic generators, $X_a(z)$ and $Y^a(z)$, with OPE
\begin{equation}
X_a(z) Y^b(w) \sim \frac{\delta_a^b}{z-w}
\end{equation}
and all other OPE trivial. All fields have conformal dimension $1/2$.

The stress tensor can be taken to be 
\begin{equation}
T = \frac{1}{2} X_a \partial Y^a - \frac{1}{2} Y^a \partial X_a
\end{equation}
with central charge $-N$. 

The current algebra contains a WZW $Sp(N)_{-\frac{1}{2}}$ current subalgebra: 
\begin{equation}
J^-_{ab} = \frac{1}{2} X_a X_b \qquad \qquad J_a^b =\frac{1}{2} X_a Y^b \qquad \qquad J_+^{ab} = \frac{1}{2} Y^a Y^b
\end{equation}
Here $J_a^b$ are the currents for an $U(N)_{-1}$ current subalgebra and $J^-_{ab}$, $J_+^{ab}$ the remaining currents in $Sp(N)_{-\frac{1}{2}}$.
Furthermore, $T$ coincides with the Sugawara stress tensor for $Sp(N)_{-\frac{1}{2}}$. \footnote{In detail, \begin{equation}
: J_a^b J_b^a : = \frac{1}{4} X_a X_b Y^a Y^b +\frac{N}{4} X_a \partial Y^a - \frac{N}{4} Y^a \partial X_a
\end{equation}
and 
\begin{equation}
: J^-_{ab} J_+^{ab} : = \frac{1}{4} X_a X_b Y^a Y^b +\frac{N+1}{2} Y^a \partial X_a \qquad : J_+^{ab} J^-_{ab} : = \frac{1}{4} X_a X_b Y^a Y^b - \frac{N+1}{2} X_a \partial Y^a
\end{equation}
so that 
\begin{equation}
: J_a^b J_b^a : -\frac{1}{2}: J^-_{ab} J_+^{ab} : - \frac{1}{2}: J_+^{ab} J^-_{ab} : = (N+\frac{1}{2}) T
\end{equation}
}The central charge matches as well.
Thus $\Sb[\C^{2N}]$ can be interpreted as an extension of an $Sp(N)_{-\frac{1}{2}}$ VOA. The $(X_a, Y^a)$ fields can be identified with the $Sp(N)_{-\frac{1}{2}}$ primaries in the fundamental representation. 

We can also focus on the $U(N)_{-1}$ currents
Notice also that 
\begin{equation}
: J_a^a J_b^b : = \frac{1}{4} X_a X_b Y^a Y^b +\frac{1}{4} X_a \partial Y^a - \frac{1}{4} Y^a \partial X_a
\end{equation}
and thus 
\begin{equation}
 T  = \frac{2}{N-1} : J_a^b J_b^a : -  \frac{2}{N-1} : J_a^a J_b^b : = T_{SU(N)_{-1}} + T_{U(1)}
\end{equation}
Thus $T$ also coincides with the Sugawara stress tensor for $U(N)_{-1}$.The central charge matches as well.

The $X_a$ and $Y^a$ fields can be identified with the $U(N)_{-1}$ primaries in the fundamental or anti-fundamental representation. 
Notice that the dimension $1/2$ receives a contribution $1/2 + 1/(2N)$ from $SU(N)_{-1}$ and $-1/(2N)$ from $U(1)$. 

The vertex algebra of $N$ symplectic bosons should contain infinitely many $U(N)_{-1}$ primaries. For example, the symmetric polynomials 
$X_{a_1} \cdots X_{a_n}$ should be $U(N)_{-1}$ primaries labelled by the symmetric powers of the fundamental representation, and 
$Y^{a_1} \cdots Y^{a_n}$ should be $U(N)_{-1}$ primaries labelled by the symmetric powers of the anti-fundamental representation.
There may be other primaries as well, hidden deeper into the symplectic bosons Verma module. 

The current algebras we identified imply that the VOA of $N$ symplectic bosons will give D-modules on $\Bun(Sp(N),C)$ (or better,
the modification of that which parameterizes bundles of the form $E_{Sp(N)} \otimes K^{\frac12}$) or on $\Bun(U(N),C)$ 
(or better, the modification of that which parameterizes bundles of the form $E_{U(N)} \otimes K^{\frac12}$).

These D-modules encode the BAA branes associated to certain boundary conditions for the corresponding four-dimensional gauge theories. 

The S-dual of these boundary conditions is known. For the $Sp(N)'$ boundary condition (here the prime indicates the presence of a discrete $\theta$ angle, 
which makes the $Sp(N)'$ theory self-S-dual) that is a maximal Nahm pole. For the $U(N)$ boundary condition that is a sub-regular Nahm pole, 
breaking the gauge group to a $U(1)$ subgroup, which is gauged at the boundary. The BBB image of Nahm pole boundary conditions is poorly understood, though. 
It would be very interesting to test this expectation, say by computing Hecke modifications of the symplectic bosons D-module. 

Another setup involving fundamental hypermultiplets is that of a D5 interface between two $U(N)$ theories or a half-D5 between an $Sp(N)$ and an 
$Sp(N)'$ theory. The BAA image of that is a D-module on the product of two copies, say, of $\Bun(U(N),C)$ localized on the diagonal. It can also be interpreted as a functor 
mapping D-modules on, say, $\Bun(U(N),C)$ to D-modules on the same space. The functor consists simply of taking a tensor product with 
the D-module defined by the VOA of $N$ symplectic bosons. 

The S-dual of (half-)D5 interfaces are (half-)NS5 interfaces, which we will discuss momentarily \cite{Gaiotto:2008sa,Gaiotto:2008ak}.

More general D5 interfaces can be defined between $U(N)$ and $U(M)$ gauge groups with different ranks $M<N$
(or symplectic groups or orthogonal with different ranks) but they are simpler and do not involve boundary degrees of freedom, only certain Nahm poles.  
They give D-modules on the product of $\Bun(U(N),C)$ and $\Bun(U(M),C)$ localized on the image of the block-diagonal embedding 
of $U(M)$ bundles into $U(N)$ bundles. 

\subsection{Bi-fundamental free hypermultiplets, H-twist}
We now organize $N M$ symplectic bosons into two $N \times M$ blocks, $X^i_a(z)$ and $Y_i^a(z)$, with OPE
\begin{equation}
X^i_a(z) Y_j^b(w) \sim \frac{\delta_a^b \delta^i_j}{z-w}
\end{equation}
and all other OPE trivial. All fields have conformal dimension $1/2$.

The stress tensor can be taken to be 
\begin{equation}
T = \frac{1}{2} X^i_a \partial Y_i^a - \frac{1}{2} Y_i^a \partial X^i_a
\end{equation}
with central charge $-N M$. 

We can define $SU(N)_{-M} \times SU(M)_{-N} \times U(1)$ currents
\begin{equation}
J_a^b =\frac{1}{2} X^i_a Y_i^b -\frac{\delta_a^b}{2N}X^i_c Y_i^c \qquad  J_j^i =\frac{1}{2} X^i_a Y_j^a -\frac{\delta^i_j}{2N}X^k_a Y_k^a \qquad J = \frac{1}{2} X^i_a Y_i^a
\end{equation}
and denote for convenience as $\tilde J_a^b$ and $\tilde J_j^i$ the bilinear currents without traces removed, which generate $U(N)_{-M}$ and $U(M)_{-N}$ 
current algebras.

We can compute
\begin{equation}
: \tilde J_a^b \tilde J_b^a : = \frac{1}{4} X^i_a X^j_b Y_j^a Y_i^b +\frac{N}{4} X^i_a \partial Y_i^a - \frac{N}{4} Y_i^a \partial X^i_a
\end{equation}
and 
\begin{equation}
: \tilde J_j^i \tilde J_i^j : = \frac{1}{4} X^i_a X^j_b Y_j^a Y_i^b +\frac{M}{4} X^i_a \partial Y_i^a - \frac{M}{4} Y_i^a \partial X^i_a
\end{equation}
and thus 
\begin{equation}
 T  = \frac{2}{N-M} : \tilde J_a^b \tilde J_b^a : -  \frac{2}{N-M} : \tilde J_j^i \tilde J_i^j : = T_{SU(N)_{-M}} +T_{SU(M)_{-N}} + T_{U(1)}
\end{equation}
Thus $T$ also coincides with the Sugawara stress tensor for the $SU(N)_{-M} \times SU(M)_{-N} \times U(1)$ currents. The central charge matches as well. This is a non-unitary analogue of level-rank duality. 

The $X^i_a$ and $Y_i^a$ fields can be identified with the $SU(N)_{-M} \times SU(M)_{-N} \times U(1)$  primaries in the bi-fundamental representation. 
The symmetric polynomials $X^{i_1}_{a_1} \cdots X^{i_n}_{a_n}$ can be decomposed into sums of products of irreducible irreps of the permutation group,
which will be primaries of $SU(N)_{-M} \times SU(M)_{-N} \times U(1)$ in the corresponding representations, and so on. 

In a similar manner, the $N \times M$ hypermultiplets can be re-organized in terms of $Sp(N)_{-\frac{M}{2}}$ and $SO(M)_{-N}$ WZW current sub-algebra,
i.e. transforming as the ortho-symplectic version of a bi-fundamental field. We will discuss the $N=1$ case momentarily.

Bi-fundamental hypermultiplets are the basic building blocks for NS5 and half-NS5 interfaces. They can give us D-modules on products of 
spaces of bundles or functors mapping D-modules on a space of bundles to another. These BAA objects will be dual to the BBB objects built from 
D5 and half-D5 interfaces. 

\subsection{$N$ free hypermultiplets as $SU(2)$ doublets, H-twist}
It is interesting to take $N$ copies of the symplectic boson and look at the properties of the 
$SU(2)_{-N/2}$ current algebra which acts diagonally on them. We can organize the fields into 
 $SO(N)$ fundamentals:
\begin{equation}
X_i(z) Y_j(w) \sim \frac{\delta_{ij}}{z-w}
\end{equation}
The $SU(2)_{-N/2}$ currents take the form 
\begin{equation}
J^- = \frac{1}{2} X_i X_i \qquad \qquad J^3 =\frac{1}{2} X_i Y_i \qquad \qquad J^+ = \frac{1}{2} Y_i Y_i
\end{equation}

We also have an $SO(N)_{-2}$ current algebra (we normalize them in the same way as a level $1$ currents $\psi_i \psi_j$ in a theory of $N$ free fermions). 
\begin{equation}
J_{ij} =\frac{1}{2} X_i Y_j - \frac{1}{2} X_j Y_i
\end{equation}
The central charge at such level is $c_{SO(N)_{-2}} = -\frac{N(N-1)}{N-4}$, which combines with the central charge of the $SU(2)$ algebra $c_{SU(2)_{-N/2}} =  \frac{3N}{N-4}$ to give the total 
central charge $-N$ of the symplectic bosons. The total stress tensor is actually the sum of the Sugawara stress tensors for the two current algebras. 
\footnote{
Indeed
\begin{equation}
: J^3 J^3 : = \frac{1}{4} (X \cdot Y)^2 +\frac{1}{4} X \cdot \partial Y - \frac{1}{4} Y \cdot \partial X
\end{equation}
and 
\begin{equation}
: J^- J^+ : = \frac{1}{4} X^2 Y^2 +Y \cdot \partial X \qquad : J^+ J^- : = \frac{1}{4} X^2 Y^2 - X \cdot \partial Y
\end{equation}
On the other hand, 
\begin{equation}
: J_{ij} J^{ij} : = \frac{1}{2} X^2 Y^2 - \frac{1}{2} (X \cdot Y)^2- \frac{N-1}{2} X \cdot \partial Y + \frac{N-1}{2} Y \cdot \partial X
\end{equation}
so that 
\begin{equation}
: J^3 J^3 : -\frac{1}{2}: J^- J^+ : - \frac{1}{2}: J^+ J^- : +\frac{1}{2} : J_{ij} J^{ij} := \frac{4-N}{2} T
\end{equation}}

\subsection{$C$-twist of a single hypermultiplet}
The vertex algebra $\Fc[\C^2]$ has two fermionic generators, $x(z)$ and $y(z)$, with OPE
\begin{equation}
x(z) y(w) \sim \frac{1}{(z-w)^2}
\end{equation}
and conformal dimension $1$. We can denote them as ``fermionic currents''. They can be also thought of 
a $PSU(1|1)$ current algebra. 

The stress tensor can be taken to be 
\begin{equation}
T = - x y
\end{equation}
and gives $c_{xy}=-2$. \footnote{
Indeed, we have OPE
\begin{equation}
T(z) x(w) \sim \frac{x(w)}{(z-w)^2} + \frac{\partial x(w)}{(z-w)} \qquad T(z) y(w) \sim \frac{y(w)}{(z-w)^2} + \frac{\partial y(w)}{(z-w)}
\end{equation}
and 
\begin{equation}
T(z) T(w) \sim -\frac{1}{(z-w)^4} +  \frac{2 T(w)}{(z-w)^2} + \frac{\partial T(w)}{z-w}
\end{equation}}

This vertex algebra can be found in several free CFTs, but in these realizations either $x$ or $y$ or both 
are derivatives of some dimension $0$ operator. These realizations clearly produce some (sections of the sheaf of) conformal blocks, 
but not necessarily all of them.

\subsubsection{Coupling to flat bundles}
Although the algebra has an $SU(2)_o$ outer automorphisms rotating $x$ and $y$, it has no corresponding current algebra. 
When working in an $SU(2)_o$ covariant way, we can denote the currents as $z_\alpha$. 

The vertex algebra can be coupled to an $SU(2)_o$ complexified local system and the dependence on the holomorphic part 
of the connection will not drop out. We expect conformal blocks to define a sheaf on $\Loc(SU(2),C)$.
(We denote the group with the compact form, but we refer to local systems for the complexified group). 

It is convenient to represent the local system as a D-module, i.e. prescribe an $SU(2)_o$ bundle $E$ equipped with an holomorphic 
$SU(2)_o$ connection $A_{\alpha \beta}(z)$. The connection modifies the OPE to  
\begin{equation}
z_\alpha(z) z_\beta(w) \sim \frac{\epsilon_{\alpha \beta}}{(z-w)^2} + \frac{A_{\alpha \beta}}{z-w}
\end{equation}
In order to find conformal blocks we need to solve the Ward identities associated to these OPE 
with currents which are sections of $E$ on $C$. 

The space of conformal blocks for a generic local system has dimension $2^{2g-2}$ and can be identified with the Fock space built from 
the $(2g-2)$ holomorphic sections $\omega^a_\alpha(z)$ of $E \otimes K$. Essentially, we can postulate that 
correlation functions with less than $n$ insertion vanish and correlation functions with exactly $n$ insertions are
\begin{equation}
\langle Z_{\alpha_1}(z_1) \cdots Z_{\alpha_n}(z_n)\rangle \sim \omega^{[a_1}_{\alpha_1}(z_1)\cdots \omega^{a_n]}_{\alpha_n}(z_n)
\end{equation}
Other correlation functions are determined from the Ward identities. 

\subsection{Free hypermultiplets in fundamental or bi-fundamental representations, C-twist}
A collection of $N$ fermionic currents $x^a(z)$ and $y_a(z)$, with OPE
\begin{equation}
x^a(z) y_b(w) \sim \frac{\delta^a_b}{(z-w)^2}
\end{equation}
give a VOA $\Fc[\C^{2N}$ with an $Sp(N)_o$ group of outer automorphisms, with an obvious $U(N)_o$ subgroup. 
It can be coupled to an $Sp(N)$ or $U(N)$ bundles equipped with a holomorphic connection. 

A collection of $N \times M$ fermionic currents $x_i^a(z)$ and $y^i_a(z)$, with OPE
\begin{equation}
x^a_i(z) y_b^j(w) \sim \frac{\delta^a_b \delta_i^j}{(z-w)^2}
\end{equation}
has an obvious action of $U(N)\times U(M)$. It can be coupled to $U(N) \times U(M)$ bundles equipped with a holomorphic connection. 
Similar considerations apply to $Sp(N) \times SO(M)$ actions. 

These VOA will appear when one studies the BBB images of D5 and NS5 interfaces.  

\section{Abelian examples}
Mirror symmetry is well understood for Abelian gauge theories. This provides us with 
some important checks of our proposal. 

Notably, a $U(1)$ gauge theory (SQED) coupled to a single hypermultiplet of charge $1$ is mirror to a single 
free hypermultiplet. All other Abelian mirror symmetries follow from repeated applications of this simple duality relationship. Another important example is SQED coupled to two hypermultiplets of charge $1$, which gives a UV description of $T[SU(2)]$ and is self-mirror. 

Furthermore, S-duality for a 4d $U(1)$ gauge theory acts in a very simple manner on boundary degrees of freedom: a 3d theory $T$ with a $U(1)$ factor in $G_H$ considered 
as a boundary condition for a 4d $U(1)$ gauge theory is mapped to a S-dual theory $T'$ obtained from $T$ by gauging the $U(1)$. 
The theory $T'$ has an obvious $U(1)_C$ factor in $G_C$. Applying mirror symmetry to $T'$ 
we obtain a new theory ${}^\vee T$ with a $U(1)_H$ factor in $G_H$, the S-dual to $T$. 

If $T$ is associated to a VOA $\bA$, then $T'$ is associated to the coset $\frac{\bA}{U(1)}$. 
It should be possible to argue in general that the conformal blocks for $\bA$ and $\frac{\bA}{U(1)}$
give Geometric Langland dual objects for a $U(1)$ gauge group, by matching the 
coset construction (possibly in the BRTS formalism) with an appropriate Fourier-Mukai 
transformation.  

\subsection{SQED with one flavor, H-twist}
Following our prescription, we need to take the coset of the $XY$ system by the $U(1)$ current 
algebra generated by $J^3$. The coset will be endowed with a $G_C = U(1)_o$ global symmetry.
If our prescription is correct, we should obtain the same VOA as in the C-twist of a single free hypermultiplet. 

Taking cosets by Abelian current algebras is a relatively simple procedure: we take primary operators of charge $q$ 
under $J_3$ and strip off a $U(1)$ vertex operator of charge $q$. For the $XY$ model, this is essentially the standard bosonization 
of a $\beta \gamma$ system: we write 
\begin{equation}
J^3 = \frac{1}{2} \partial \phi \qquad X = e^{- \phi} x \quad Y = e^\phi y
\end{equation}

The notation $x$ and $y$ is completely intentional: $x$ and $y$ are fermionic currents of conformal dimension $1$, charge $\pm 1$ under $U(1)_o$
and free OPE 
\begin{equation}
x(z) y(w) \sim \frac{1}{(z-w)^2}
\end{equation}
Notice that the central charges match: $c_{XY} = c_{xy} + c_{J^3} = -2+1$. 

Taking the coset of the $XY$ system by the algebra generated by $J_3$ leaves us with the 
algebra of $x$ and $y$. This is beautifully consistent with the mirror symmetry relation between SQED with one flavor and 
a theory of a free hypermultiplet. Notice that the $U(1)_o$ global symmetry of the coset coincides with the $U(1)_o$ global symmetry 
of the $x$ and $y$ fermionic currents. 
The bosonization relation can be stated as
\begin{equation}
\Fc[\C^2] = \frac{\Sb[\C^2]}{\hat U(1)_{-1}}
\end{equation}
i.e. 
\begin{equation}
\bA_H[\mathrm{SQED}_1] = \bA_C[\mathrm{Free\,\,hyper}]
\end{equation}

It is also straightforward, but rather non-trivial, to verify that the characters of the vacuum module for the symplectic bosons decomposes appropriately into 
free bosons characters and characters for the fermionic currents:
We can expand 
\begin{equation}
\chi^{XY}  = \frac{1}{\prod_{n=0}^\infty (1- q^{n+1})^2}\sum_{n=0}^\infty \sum_{m=-n}^n z^m (-1)^{n-m} q^{\frac{n(n+1)-m^2}{2}}
\end{equation}
The replacement
\begin{equation}
\chi^{U(1)}_m(q,z)= \frac{z^m q^{-\frac{m^2}{2}}}{\prod_{n=0}^\infty (1- q^{n+1})} \to t^m
\end{equation}
corresponds to stripping off the free boson Verma module, while keeping track of the $U(1)_o$ charge. It would map 
\begin{equation}
\chi^{XY}  \to \frac{1}{\prod_{n=0}^\infty (1- q^{n+1})} \sum_{n=0}^\infty q^{\frac{n(n+1)}{2}} \frac{t^{n+1/2}+t^{-n-1/2}}{t^{1/2} + t^{-1/2}}
\end{equation}
which can be rewritten as 
\begin{equation}
\chi^{XY}  \to \frac{1}{\prod_{n=0}^\infty (1- q^{n+1})} \frac{1}{t^{1/2} + t^{-1/2}}\sum_{n=-\infty}^\infty q^{\frac{n(n+1)}{2}} t^{n+1/2}
\end{equation}
and then 
\begin{equation}
\chi^{XY} \to \prod_{n=0}^\infty (1- t q^{n+1})(1- t^{-1} q^{n+1})
\end{equation}

The right hand side id the character for the fermionic current algebra, graded by the $U(1)_o$ charge. 
Thus we can write 
\begin{equation}
\chi^{XY} = \sum_{n=-\infty}^\infty \chi^{U(1)}_n \chi^{xy}_n
\end{equation}
where $\chi^{xy}_n$ is the charge $n$ part of the character of the $xy$ fermionic current algebra. 
\footnote{The bosonization relation between $X,Y$ and $x,y$ is simple, but already for correlation functions on the sphere 
it leads to intricate identities between rational functions. As an example, we can look at a four-point function:
\begin{equation}
\langle X(z) X(z') Y(w) Y(w') \rangle = \frac{1}{(z-w)(z'-w')} + \frac{1}{(z-w')(z'-w)}
\end{equation}
The $U(1)$ part of the correlation function is $\frac{(z-w)(z-w')(z'-w)(z'-w')}{(z-z')(w-w')}$. Stripping it off we get the rational function
\begin{equation}
\frac{(z-z')(w-w')}{(z-w)^2(z'-w')^2(z-w')(z'-w)} + \frac{(z-z')(w-w')}{(z-w)(z'-w')(z-w')^2(z'-w)^2}
\end{equation}
which can be reorganized to 
\begin{equation}
\langle x(z) x(z') y(w) y(w') \rangle = \frac{1}{(z-w')^2(z'-w)^2}-\frac{1}{(z-w)^2(z'-w')^2}
\end{equation}}

\subsubsection{Twisted sector}
We have observed that generic $U(1)$ twisted sectors $|\alpha\rangle$ for the symplectic bosons have a $U(1)$ charge proportional to the $U(1)$ twist. 
Indeed, it is well-known that such twisted sectors can be bosonized to the basic vertex operators $e^{\alpha \phi}$. The conformal dimensions match 
and these vertex operators induce the correct monodromy in $e^{\pm \phi}$. In particular, they are mapped back to the 
vacuum module under bosonization. 

In order to find twisted sectors for the fermionic current VOA, we need to look at the $|R,\lambda \rangle$ general Ramond modules for the 
symplectic bosons. Indeed, these have $U(1)$ charge $\lambda - \frac12$ and the corresponding $U(1)$ vertex operator 
$e^{(\lambda - \frac12)\phi}$ would induce a monodromy $- e^{\pm 2 \pi \lambda}$ on the $e^{\pm \phi}$ vertex operators
which appear in the symplectic bosons. Thus the $|R,\lambda \rangle$ general Ramond modules should contain a 
twisted sector with monodromy $e^{\pm 2 \pi \lambda}$ for the fermionic currents.

Because of the role the $|R,\lambda \rangle$ general Ramond modules play in defining 
the hidden conformal blocks of the symplectic boson VOA, this fact also suggests that the hidden blocks 
should bear some relationship to the conformal blocks for the fermionic currents VOA coupled to general 
$U(1)$ local systems.

\subsubsection{Coset vs BRST reduction}
If we are given a vertex algebra which has a level $0$ $U(1)$ current $J$, a nice BRST construction becomes available: 
we can add a system of $bc$ ghosts of dimensions $(1,0)$ and define the BRST charge
\begin{equation}
Q = \oint c J
\end{equation}
This will have the effect of removing from the theory $J$ and all operators which are charged under $J$.
Operators in the original VOA which have trivial OPE with $J$ will remain as BRST-closed operators. 

In our setup, the symplectic bosons have a $U(1)$ current of level $-1$. We can add to them a standard system of 
complex fermions 
\begin{equation}
\psi(z) \chi(w) \sim \frac{1}{z-w}
\end{equation}
which have a single conformal block and a $U(1)$ current at level $1$ and then take the BRST reduction
with respect to $J = X Y + \psi \chi$. 

The result of this BRST reduction appears to be the same as the coset we discuss in this section, 
including the sectors of non-trivial $U(1)_o$ charge: 
operators of charge $n$ in the theory of symplectic bosons can be dressed with charge $n$ primaries of the free fermion VOA in order to give 
BRST cohomology classes, which have the same dimension and properties as the corresponding operators in the coset. 

For example, we would identify the basic BRSt closed operators with the fermionic currents
\begin{equation}
X(z)\psi(z) \to x(z) \qquad \qquad Y(z) \chi(z) \to y(z) 
\end{equation}
The BRST reduction is a bit more systematic than the coset. In particular, it gives a more precise way to built 
conformal blocks of the coset theory, rather than attempting an expansion of symplectic boson correlation functions into 
a product of $U(1)_{-1}$ conformal blocks and coset blocks. 

Notice the OPE
\begin{equation}
X(z) \psi(z) Y(w)\chi(w) \sim \frac{1}{(z-w)^2}+ \frac{J(w)}{z-w}
\end{equation}
As the $U(1)_o$ symmetry is identified with the global symmetry of the $\psi$ and $\chi$ fermions, we can couple the system naturally to an 
$U(1)_o$ bundle by coupling the fermions themselves. Furthermore, we can add a coupling to an $U(1)_o$ holomorphic connection $A_0(z)$ by deforming the BRST 
operator to 
\begin{equation}
Q = \oint dz c(z) \left(J(z) - A_o(z)\right)
\end{equation}
This allows us to couple the coset theory to a full $U(1)_o$ local system. 

The $bc$ system has an anomaly, which forces us to introduce at least $g$ insertions of $b(z)$ and 
one insertion of $c(z)$. The corresponding correlation function is
\begin{equation} \label{eq:bc}
\langle b(z_1) \cdots b(z_g) c(w) \rangle \sim \det_{ij} \omega^{a_i}(z_j)
\end{equation}
where $\omega^{a_i}(z_j)$ is a basis of holomorphic differentials. 

The $b(z)$ insertions are not BRST closed! Rather, 
\begin{equation}
\{Q, b(z) \} = J(z) - A_o(z)
\end{equation}
If we contract the $b(z)$ insertion with an anti-holomorphic differential $\delta \bar A$
then the first term in the right hand side $\int dz d\bar z J(z)\delta \bar A$
is a total derivative along $Bun(U(1),C)$. 

In the absence of $A_o(z)$, that means that we can identify the correlation functions of the combined 
system of symplectic bosons, complex fermions and ghosts as a top holomorphic form in $\Bun(U(1),C)$, 
mapped to an exact form by $Q$. Integrating the correlation function over a middle-dimensional cycle $\Bun(U(1),C)$
gives a BRST-invariant answer, which we plan to identify with a conformal block for the 
coset theory, the fermionic currents:
\begin{align} \label{eq:BRSTcoset}
&\langle x(z_1) \cdots y(w_1) \cdots   \rangle_{\Fc[\C^2],\bar A_0;\Gamma} = \cr &\oint_{\Gamma \in \Bun(U(1),C)} D\bar A \langle X(z_1) \cdots  Y(w_1) \cdots  \rangle_{\Sb[\C^2],\bar A} \langle \psi(z_1) \cdots  \chi(w_1) \cdots   \rangle_{\psi \chi,\bar A+ \bar A_o} 
\end{align}
where the measure $D\bar A$ is given by the $bc$ system correlation function \ref{eq:bc}.

In the presence of $A_o(z)$, the BRST transformation of the measure 
involves an extra constant 1-form $\int dz d\bar z A_o(z)\delta \bar A$ on the $\Bun(U(1),C)$ torus.
That means that the correlation function becomes BRST closed when the measure \ref{eq:bc} is multiplied by an appropriate Fourier 
kernel $e^{S[A_o, \bar A]}$ such that $\frac{\delta S}{\delta \bar A} = A_o(z)$. 

This seems a rather reasonable way to do a Fourier-Mukay-like transformation mapping
the D-module of conformal blocks for symplectic bosons to the sheaf of conformal blocks for the 
fermionic currents. This should be a direct manifestation of the fact that S-duality for $U(1)$ gauge theory maps 
the boundary condition associated to a single free hypermultiplet back to itself. 

It would be nice to mimic in this setup the classical mirror symmetry relationship described in \cite{Gaiotto:2016aa}.
A crucial role there was played by the $2g-2$ points on the surface where the $U(1)$ Higgs field vanishes
and by the $2^{2g-2}$ ways they could be distributed between the $X$ and $Y$ classical sections. 
 
The symplectic boson theory has conformal blocks which are localized on 
the $\Theta_d$ locus in $\Bun_d(U(1),C)$ where zeromodes appear, which is parameterized by the $g-1-d$ zeroes of 
the negative charge zeromode for positive $d$ and by the $g-1+d$ zeroes of 
the positive charge zeromode for non-positive $d$. 

We expect these conformal blocks to map to the conformal blocks of fermionic currents which have a $U(1)_o$ anomaly 
$d$. A possible explanation would be that for a general point on $\Loc(C,U(1))$ the integrand of \ref{eq:BRSTcoset} is not single-valued on $\Bun(U(1),C)$.
Good integration cycles would consist of a small loop around $\Theta_d$ times a contour integral over $\Theta_d$. It would be nice to give a detailed derivation of this relationship.
  
We can give a toy demonstration of this for a torus partition function. The contour integral for a partition function is 
\begin{equation}
\oint da \eta^2(\tau) e^{2 \pi i a b_0} \frac{\eta(\tau)}{\theta(a,\tau)}\frac{\theta(a+a_o,\tau)}{\eta(\tau)}
\end{equation}  
Here the first factor is the $bc$ partition function, followed by the Fourier kernel, the $XY$ partition function and the $\psi \chi$ partition functions. 

If we take the contour to run around the pole at $a=0$ we get 
\begin{equation}
\oint_0 da \eta^2(\tau) e^{2 \pi i a b_0} \frac{\eta(\tau)}{\theta(a,\tau)}\frac{\theta(a+a_o,\tau)}{\eta(\tau)} = \frac{\theta(a_0,\tau)}{\eta(\tau)}
\end{equation}  
which is the character for the fermionic currents in a general $U(1)_o$ background, with no zeromodes. 

On the other hand, if $b_0$ is $0$ (or an integer $n$) we can take a contour integral on the $A$ cycle of the torus
and the calculation mimics the character computations done earlier int he section, yielding the vacuum character $\chi_{xy}(a_0, \tau)$.
Contour integrals along other cycles of the torus impose other linear constraints on $b_0$ and $a_0$ and give other modular images of 
$\chi_{xy}(a_0, \tau)$. 

\subsection{SQED with $N$ flavors, H-twist}

We start from $N$ copies of the symplectic bosons and take a coset by the diagonal $U(1)$ current. 
According to our prescription 
\begin{equation}
\bA_H[\mathrm{SQED}_N] = \frac{\Sb[\C^{2N}]}{\hat U(1)_{-N}}
\end{equation}
The coset operation will strip $U(1)$ vertex operators of appropriate charge from the symplectic bosons $X_a$ and $Y^a$.
The resulting fields $A_a$, $B^a$ can be thought as $SU(N)_{-1}$ primaries in the fundamental or anti-fundamental representation,
of conformal dimension $1/2 + 1/(2N)$ and charge $\pm 1$ under $U(1)_o$. 

More generally, symmetric polynomials of $X_a$ or symmetric polynomials of $Y^a$
will become WZW primaries of conformal dimension $n/2 + n^2/(2N)$ with charge $\pm n$ under $U(1)_o$. 

The coset vertex algebra contains an $SU(N)_{-1}$ current algebra and the stress tensor is the Sugawara stress tensor. 
The coset current algebra, though, is larger than $SU(N)_{-1}$. It includes, for example, the fields which arises from the $N$-th symmetric powers of 
$X_a$ and $Y_a$, which have dimension $N$ and $U(1)_o$ charge $1$. It also includes operators with no $U(1)_o$ charge, such as 
an operator $O^b_a$ of dimension $2$ in the adjoint of $SU(N)_{-1}$, built by removing the $U(1)$ contributions from $X_a \partial Y^b - Y^b \partial X_a$.

As the dimension $3$ operators with no $U(1)_o$ charge seem to be accounted fully by currents or derivatives of  $O^b_a$, it would appear that the 
$SU(N)_{-1}$ currents, together with $O^b_a$, have a closed set of OPE, with the $OO$ OPE 
involving bilinears and derivatives of the currents themselves. The existence of such a self-consistent, associative OPE is typically rather non-trivial
and could be taken perhaps as an alternative definition of the subsector with no $U(1)_o$ charge in the coset VOA. 

It is interesting to mimic the basic trick of Abelian mirror symmetry: apply the basic mirror symmetry operation to all hypermultiplets.  
We can apply the bosonization formula to each symplectic boson: 
\begin{equation}
X_a Y^a = \partial \phi_a \qquad X_a = e^{- \phi_a} x_a \quad Y^a = e^{\phi_a} y^a
\end{equation}
This bosonization hides the $U(N)_{-1}$ current algebra, leaving only a Cartan subalgebra manifest. 

The coset by the diagonal $U(1)$ current is now elementary: we simply impose the constraint $\sum_a \phi_a =0$.
This gives a free field realization of the coset vertex algebra in terms of $\Fc[\C^{2N}]$ and $N-1$ free bosons. 

The $N=2$ example is rather special and deserves a separate treatment. We will come back to that momentarily. 

Finally, we can consider a BRST reduction which should reproduce the coset VOA: 
we couple the $N$ symplectic bosons to a $U(1)_N$ current algebra in order to get a level $0$ current $J$.
Adding a set of $bc$ ghosts gives us a BRST current $c J$ and we can pass to BRST cohomology. Again, the 
extra $U(1)$ current, the ghosts and the $U(1)_{-N}$ currents will cancel out and leave the coset fields behind. 

The conformal blocks for the coset VOA will define a D-module on $\Bun(SU(N),C)$
which is also a sheaf on $\Loc(U(1)_o,C)$. If this theory is used to define a boundary condition for a 4d $SU(N)$ gauge theory, 
the result is the S-dual of the sub-regular Nahm pole boundary condition breaking the gauge group to a $U(1)$ subgroup,
which has Dirichlet b.c. and whose background connection is the point in $\Loc(U(1)_o,C)$. This statement should have a direct 
interpretation in terms of Hecke modifications of the D-module on $\Bun(SU(N),C)$. It would be nice to understand this better. 
 
\subsection{The $T[SU(2)]$ theory, H-twist}
This is a self-mirror theory with an enhancement to $SU(2)_o$ of the naive $U(1)_o$ tri-holomorphic isometry of the Coulomb branch.
We would like to see if the symmetry enhancement is manifest in our proposed current algebra. If so, this will be a strong test of our proposal. 
Furthermore, this theory plays a role in S-duality for four-dimensional $\CN=4$ SYM with 
an $SU(2)$ gauge group. The corresponding vertex algebra should be a duality kernel in Geometric Langlands 
for the group $SU(2)$. 

A computation of the character for the coset of two symplectic bosons by the diagonal $U(1)$ 
current algebra produces a very pleasing result: 
\begin{equation}
\chi_{\bA_H[T[SU(2)]]} = \sum_{j=0}^{\infty}(t^j + t^{j-1} + \cdots + t^{-j})\chi_{j}[SU(2)_{-1}] 
\end{equation}
where $t$ is the fugacity for the $U(1)^o$ global symmetry. 

We can recast the character as 
\begin{equation}
\chi_{\bA_H[T[SU(2)]]} = \sum_{j=0}^{\infty}\chi_j[SU(2)_o] \chi_{j}[SU(2)_{-1}] 
\end{equation}
This is a clearly compatible with the idea that the outer $U(1)$ global symmetry has been promoted to an outer $SU(2)_{o}$ 
global symmetry, in such a way that the spin $j$ primaries of the $SU(2)_{-1}$ WZW current algebra transform in a spin $j$ representation of 
$SU(2)_{o}$. The $SU(2)_o$ enhancement seems to be previously known 
to experts in the subject.

We can give a bosonized description of the coset as follows: 
\begin{align}
J &= X_a Y^a = \partial \phi \cr
X_a &= e^{-\phi} A_a \qquad Y^a = e^{\phi} B^a
\end{align}
The fields $A_a$ and $B^a$ have dimension $3/4$. The outer $SU(2)_{o}$ global symmetry will rotate 
among each other $A_a$ and $B_a \equiv \epsilon_{ab} B^b$. 

We can make the $SU(2)_{o}$ global symmetry manifest if we look at the free field realization:
\begin{align}
A_1 &= e^{-\tilde \phi} x_1 \qquad B^1 = e^{\tilde \phi} y^1 \cr
A_2 &= e^{\tilde \phi} x_2 \qquad B^2 = e^{-\tilde \phi} y^2
\end{align}
The $SU(2)_{o}$ global symmetry is a subgroup of the automorphism group of the four fermionic currents $x_a$, $y^a$, 
which rotates the doublets $(x_1, y^2)$ and $(x_2, -y^1)$.  

In a better notation, we can write the parameterization as 
\begin{align}
C^\alpha_1 &= e^{-\tilde \phi} z^\alpha_1 \qquad C^\alpha_2 = e^{\tilde \phi} z^\alpha_2 
\end{align}
with OPE 
\begin{equation}
z^\alpha_1(z) z^\beta_2(w) \sim \frac{\epsilon^{\alpha \beta}}{(z-w)^2}
\end{equation}
with $SU(2)_{o}$-invariant WZW currents 
\begin{align}
J^3 &= \frac{1}{2} \partial \tilde \phi \cr
J^- &= e^{-2 \tilde \phi} z^\alpha_1 z^\beta_1 \epsilon_{\alpha \beta} \cr
J^+ &= e^{2 \tilde \phi} z^\alpha_2 z^\beta_2 \epsilon_{\alpha \beta}
\end{align}

Going back to the symplectic bosons, we can use a notation 
\begin{equation}
X_a(z) Y_b(w) \sim \frac{\epsilon_{ab}}{z-w}
\end{equation}
The coset current algebra includes $SU(2)_{-1}$ WZW currents and also a $SU(2)_{o}$ triplet of operators of dimension $2$. The triplet consists of the 
operator $O^0_{ab}$ built from $X_{(a} \partial Y_{b)} - Y_{(a} \partial X_{b)}$ and the two operators 
\begin{equation}
O^+_{ab} = (X_a X_b) \qquad O^-_{ab} = (Y_a Y_b)
\end{equation}
where the parenthesis denotes removing the $U(1)$ primary.

Again, it seems likely that the $O^I_{ab}$ operators form a closed OPE with the current algebra fields.
The OPE is actually manifestly $SU(2)_o$ invariant \cite{shao}. This offers a potential rout to study conformal blocks 
coupled to $SU(2)_o$ local systems by algebraic methods. It would be very interesting to do so. 

Thus our proposal is that the kernel for Geometric Langlands with gauge group $SU(2)$:
\begin{equation}
\bA_H[T[SU(2)]] = \frac{\Sb[\C^{4}]}{\hat U(1)_{-2}}
\end{equation}
\subsection{The $A_{k-1}$ theory, H-twist}
Another natural example to consider is the linear quiver of $k-1$ $U(1)$ nodes, with a single flavor at each end. 
This theory is expected to have an enhancement of the Coulomb branch symmetry from $U(1)^{k-1}_C$ to $SU(k)_C$: it is the mirror of SQED with $k$ flavors. 

If we use bosonization of the $k$ symplectic bosons, the enhancement is automatic: 
the coset removes all $U(1)$ currents except for the diagonal combination, leaving behind 
\begin{equation}
A_a = e^{- \phi} x_a \quad B^a = e^{\phi} y^a
\end{equation}
with $SU(k)_o$ global symmetry and $U(1)$ current algebra, as expected

\subsection{The $C$-twist of general Abelian gauge theories}
Abelian mirror symmetry is rather well understood and reduces to the basic mirror symmetry of SQED with one flavor. 

If we start from a general Abelian theory, build the mirror Abelian theory and $H$-twist it, we should have the same result as if we $C$-twist the original theory. 
Furthermore, if we bosonize all the symplectic bosons in the mirror theory, the bosonized coset algebra can be expressed directly in terms of the original 
Abelian theory: we take a set of fermionic currents $x_a$, $y_a$ for each of the original hypermultiplets,
add a free boson for each factor in the gauge group and dress the fermionic currents with 
gauge symmetry $q$ by a free boson vertex operator of charge $q$. 

The resulting candidate $\bA_C[T]$ vertex algebra has an Abelian current algebra generated by the free bosons, which we identify with the current algebra for $G_C$, 
and a global symmetry algebra $G_H$. 

It would be very interesting to find a non-Abelian generalization of this construction. Perhaps one can associate WZW currents to 
vectormultiplets and build some VOA by dressing WZW primary fields with the fermionic currents associated to the matter hypermultiplets.

\section{Non-Abelian examples: $SU(2)$ gauge theories}

\subsection{The $SU(2)$ SQCD theory with four flavors ($G_H=SO(8)$), H-twist}
This theory is a building block for several other examples in this paper. 
It is the dimensional reduction of a four-dimensional superconformal theory, a fact which allows us do some comparisons with the four-dimensional chiral algebras. 
It also inherits from four dimensions a non-trivial triality symmetry: in the IR, it becomes invariant under discrete symmetry transformations which act as triality on $SO(8)_H$
This symmetry is not a manifest symmetry in the UV gauge theory description. 

The coset vertex algebra must consist of a collection of modules for the $SO(8)_{-2}$ current algebra. 
Remember that the coset stress tensor coincides with the Sugawara tensor for the $SO(8)_{-2}$ current algebra.
The BRST procedure used in the 4d setup produces directly the irreducible vacuum module for the $SO(8)_{-2}$ current algebra
itself, which is triality invariant. 

In order to understand the coset procedure, we can begin to experiment with characters, decomposing the character of the symplectic bosons 
into characters for Verma modules of the $SU(2)_{-4}$ current algebra. This is likely too naive, but it is a reasonable starting point. 
We can write 
\begin{equation}
\chi_{Sb[\C^{16}]} = \sum_j \chi_{SU(2)}[V_j] \chi_j 
\end{equation}
The tentative character for the vacuum representation $\chi_0$ is ${\it not}$ invariant under triality acting on the $SO(8)$ 
fugacities. Thus it is likely not the right answer for the coset vacuum character. In particular, it does not coincide 
with the $SO(8)_{-2}$ vacuum character. 

On the other hand, something surprising happens: the combination 
\begin{equation}
\tilde \chi_0 = \chi_0 - \chi_1 + \chi_2 - \cdots
\end{equation}
appears to be triality invariant and to coincide with the $SO(8)_{-2}$ vacuum character. 
It is natural to expect this is the correct answer for the $\bA_H$ vertex algebra for this theory. 

A possible justification of this answer is that the $SU(2)_{-4}$ modules inside the symplectic bosons current algebra may be larger than Verma modules. 
For example, if the symplectic bosons current algebra involves the spin $0$ module and an extension built out of 
the spin $0$ and spin $1$ $SU(2)_{-4}$ Verma modules, the coefficient of the spin $0$ module would have to be smaller than 
the naive answer and an expression such as $\tilde \chi_0$ may appear. 

In order to seek illumination, we can look at concrete expressions for operators in the coset. 
We may start from $SU(2)$ invariant operators. Denoting the symplectic bosons as 
$Z^\alpha_i$ with $i$ being the $SO(8)$ index and $\alpha$ the $SU(2)$ index, the $SO(8)_{-2}$ 
currents are $\epsilon_{\alpha \beta} :Z^\alpha_i Z^\beta_j:$.

At level $2$, we have $\epsilon_{\alpha \beta} :Z^\alpha_i \partial Z^\beta_j:$, which includes both a symmetric and an antisymmetric 
tensors of $SO(8)$, and $\epsilon_{\alpha \beta} \epsilon_{\gamma \delta}:Z^\alpha_i Z^\beta_j Z^\gamma_k Z^\delta_t:$. The symmetric traceless tensor 
is mapped under triality to the self-dual and anti-selfdual 4-forms, which are not present in the field built from four $Z$s. Thus the symmetric traceless tensor 
at level $2$ is potentially problematic. This is also the representation which appears in the leading term in $\chi_1$, from $:Z^{(\alpha}_i Z^{\beta)}_j:$.

Acting with $SU(2)_{-4}$ current algebra operators onto $:Z^{(\alpha}_i Z^{\beta)}_j:$ we find indeed that we can produce $\epsilon_{\alpha \beta} :Z^\alpha_i \partial Z^\beta_j:$,
which is thus secretly an $SU(2)_{-4}$ descendant. this verifies the presence of indecomposable representations built from the vacuum and symmetric traceless tensor Verma modules. 

This supports the conjecture that 
\begin{equation}
\frac{\Sb[\C^{16}]}{\widehat{SU}(2)_{-4}} = \widehat{SO}(8)_{-2}
\end{equation}

\subsubsection{The $T[SU(3)]$ theory, H-twist}
This theory is a two node quiver, with $U(1) \times U(2)$ gauge group and $3$ extra flavors at the $U(2)$ gauge node. 

This is a self-mirror theory with an enhancement to $SU(3)_C$ of the naive $U(1)_C^2$ tri-holomorphic isometry of the Coulomb branch.
We would like to see if the symmetry enhancement is manifest in our proposed current algebra. If so, this will be a very strong test of our proposal. 
Furthermore, this theory plays a role in S-duality for four-dimensional $\CN=4$ SYM with 
an $SU(3)$ gauge group. The corresponding vertex algebra should be a duality kernel in Geometric Langlands 
for the group $SU(3)$. 

Conveniently, this theory is obtained by gauging two $U(1)$ symmetries of the $SU(2)$ SQCD theory with four flavors.
This will help our analysis. 
A naive computation of the character runs into the same type of trouble we encountered with this ancestor theory: 
if we decompose 
\begin{equation}
\chi_{SB} = \sum_{j,n,m} \chi_{SU(2)}[V_j] \chi^{U(1)}_n \chi^{U(1)'}_m \chi_{j,n,m} 
\end{equation}
then the naive characters 
\begin{equation}
\chi_{j} = \sum_{n,m} t_1^n t_2^m \chi_{j,n,m} 
\end{equation}
do {\it not} manifest any enhancement of the $U(1)_o^2$ naive global symmetry of the coset. 

On the other hand, if we take the same combination of naive characters as before,
\begin{equation}
\tilde \chi_0 = \chi_0 - \chi_1 + \chi_2 - \cdots
\end{equation}
we get a striking result: 
\begin{equation}
\tilde \chi_0 = \sum_\lambda \chi_\lambda[SU(3)^o](t_1,t_2) \chi_\lambda[SU(3)_{-2}]
\end{equation}
where $\chi_\lambda[SU(3)^o]$ are the characters of finite-dimensional irreps of the $SU(3)^o$ global symmetry group and 
$ \chi_\lambda[SU(3)_{-2}]$ are the characters of the Verma modules for the $SU(3)_{-2}$ current algebra. 

This is a clearly compatible with the idea that the outer $U(1)_o^2$ global symmetry has been promoted to an outer $SU(3)^{o}$ 
global symmetry, in such a way that the primaries of the $SU(3)_{-2}$ WZW current algebra transform in the corresponding representation of 
$SU(3)^{o}$. 

A simple way to build currents for the coset is to start from $SU(2)$-invariant expressions. In terms of elementary fields $X_a$, $Y_a$, $X_{i,a}$, $Y^i_a$, 
for example, we can write the three fields in the $3$ representation as
\begin{equation}
X_{i,a} X_b \epsilon^{ab} \qquad \qquad X_{i,a} Y_b \epsilon^{ab} \qquad \qquad \epsilon_{ijk}Y^j_a Y^k_b \epsilon^{ab}
\end{equation}
of dimension $4/3$.

In order to find the conjectural $SU(3)^o$ octet of adjoint currents, we need to go to level $3$. 
We can combine pairs of the fields above and their conjugate. 
Again, these octet currents should form a closed current algebra OPE. 

According to the discussion of the previous section, we can describe the coset as a $U(1)^2$ coset of the $SO(8)_{-2}$ vacuum module.
These are the Abelian factors in the subgroup $U(1) \times U(3) \subset U(4) \subset SO(8)$. The triality symmetry leaves the $SU(3)$ subgroup 
unaffected, while rotating the two $U(1)$ into each other.

This description is very invaluable, as it makes the promotion of $U(1)_o^2$ to $SU(3)_o$ evident: triality acts on the $U(1)_o^2$ charges of coset 
operators as the Weyl group, and the promotion of the $U(1)_o$ associated to the $U(1)$ node of the quiver to an $SU(2)_o$ 
was made manifest by bosonization. 

This supports the conjecture that 
\begin{equation}
\bA_H[T[SU(3)]]= \frac{\widehat{SO}(8)_{-2}}{U(1)_{-2} \otimes U(1)_{-6}}
\end{equation}
where $U(1) \otimes U(1)$ is the subgroup of $SO(8)$ which commutes with an $SU(3)$ subgroup under which the fundamental of $SO(8)$ decomposes as $8 = 3 + \bar 3 + 1 + 1$. 
\section{Non-Abelian examples: Unitary quivers} 

\subsection{$T[SU(N)]$, H-twist}
This theory is defined by a linear quiver of $U(1) \times U(2) \times \cdots \times U(N-1)$ gauge groups, with $N$ flavors for the last node. 
\begin{equation}
\bA_H[T[SU(N)]] =\frac{\Sb[\C^{2N(N-1)}] \times \Sb[\C^{2(N-1)(N-2)}] \cdots \times \Sb[\C^2]}{\hat U(N-1)_{2-2N} \cdots \hat U(1)_{-2}}
\end{equation}
The $G_H$ current algebra is thus $SU(N)_{1-N}$. Notice that all nodes have level twice the critical level for the non-Abelian gauge fields. If we remove the 
Abelian factors, the resulting theory has a four-dimensional superconformal ancestor. Let's denote the quiver without Abelian factors as $\tilde T[SU(N)]$.

Both physically and at the level of the four-dimensional chiral algebra, the $\tilde T[SU(N)]$ quiver has an enhanced symmetry which generalizes the 
triality relation we encountered for $N=2$. The enhanced symmetry acts on the $U(1)^{N-1}_H$ symmetries in the same way as the Weyl group 
acts on the Cartan torus of $SU(N)$. This is inherited from the class $S$ description. Furthermore, the four-dimensional current algebra consists of a sum or products of modules 
for the $SU(N)_{1-N}$ WZW currents tensored with modules for the $U(1)^{N-1}_H$ WZW currents.

Under the assumption that our coset operation produces the same answer as the BRST reduction which gives the four-dimensional chiral algebra, we immediately learn that 
$\bA_H[T[SU(N)]]$ has an enhanced $SU(N)_o$ global symmetry! Indeed, the triality-like global symmetry of $\bA_H[\tilde T[SU(N)]]$ 
will persist in $\bA_H[T[SU(N)]]$, acting on the $U(1)^{N-1}_o$ charges as a Weyl group. This symmetry, combined with the known symmetry enhancement at the 
$U(1)$ node of $\bA_H[T[SU(N)]]$, implies the enhancement of $U(1)^{N-1}_o$ to $SU(N)_o$.

We furthermore conjecture a character 
\begin{equation}
\chi_{\bA_H[T[SU(N)]]}= \sum_\lambda \chi_\lambda[SU(N)_o](t) \chi_\lambda[SU(N)_{1-N}]
\end{equation}

\subsection{General global symmetry enhancement}
We can now formulate a general strategy to argue that the H-twist current algebra for a general unitary 
quiver gauge theory has the expected enhancement of $G_C$.

In the UV, $G_C$ consists of a product of $U(1)$ factors, one for each unitary group in the quiver. 
At ``balanced'' nodes where the total number of flavors, including bifundamental hypers to nearby nodes and fundamental hypers to a framing node, 
equals twice the rank of the unitary group, the $U(1)$ global symmetry is enhanced to $SU(2)$. The $SU(2)$ symmetry groups 
at nearby balanced nodes combine into larger groups: $SU(k+1)$ for a chain of $k$ balanced nodes and more generally an ADE group $G_\Gamma$ 
for an ADE sub-graph $\Gamma$ of balanced nodes. 

The potential symmetry enhancement of the current algebra associated to a quiver gauge theory can also be studied 
``node by node'': we can first take the coset by the unitary group at a balanced node, and then by the remaining gauge groups. 
If the first step produces VOA with enhanced global symmetry $SU(2)_o$ commuting with the WZW symmetry 
used in the next step of the coset, that $SU(2)_o$ will persist at the next step of the calculation. 

We do not know how to demonstrate directly that the VOA for the $U(N)$ gauge theory with $2N$ flavors 
has an outer $SU(2)_o$, though the conjectural relation with 4d calculations would make the $Z_2$ Weyl group of 
$SU(2)_o$ manifest. Perhaps a direct bootstrap of the VOA from a finite set of generators demonstrating 
$SU(2)_o$ would be possible.

On the other hand, the VOA for the $U(N)$ gauge theory with $2N$ flavors can be used in the calculation of the VOA for 
$T[SU(N+k)]$ quivers. There the $U(1) \times Z_2$ symmetry of the $U(N)$ node is embedded into an $SU(2)_o$
after the coset, which involves WZW currents which commute with $U(1) \times Z_2$. This makes it at least very plausible that the 
VOA for the $U(N)$ gauge theory with $2N$ flavors does indeed have $SU(2)_o$ global symmetry. 

With that assumption, the $S_{k+1}$ Weyl group symmetry for chain of $k$ balanced nodes follow from 
the conjectural relation to the 4d calculations and combines with the $SU(2)_o$ at individual nodes to 
give the expected Coulomb branch symmetry enhancement.

\appendix

\section{D-modules and Ward identities} \label{app:dmodules}
\subsection{Examples in $T^* \C$.}
As a first toy example, consider the contour integral defining an Airy function
\begin{equation}
\Ai(x) = \oint e^{\frac{z^3}{3} + x z}  dz
\end{equation}
The differential equation for the Airy function follows from some integration by parts: 
\begin{equation}
\partial_x \Ai(x) = \oint z e^{\frac{z^3}{3} + x z}  dz \qquad \qquad \partial^2_x \Ai(x) = \oint z^2 e^{\frac{z^3}{3} + x z}  dz = - x \Ai(x)
\end{equation}
The two solutions to the differential equation can be obtained by selecting different integration contours. 
The differential equation can be cast as an holomorphic connection 
\begin{equation}
\partial_x - \begin{pmatrix} 0 & 1 \cr -x & 0\end{pmatrix}
\end{equation}

One could consider the whole vector space of correlation functions 
\begin{equation}
f_n(x) = \oint z^n e^{\frac{z^3}{3} + x z}  dz
\end{equation}
subject to the Ward identity
\begin{equation}
f_{n+2}(x) + x f_n(x)= \oint z^n \partial_z \left(e^{\frac{z^3}{3} + x z} \right) dz = - n f_{n-1}(x)
\end{equation}
with the differential acting as $\partial_x f_n(x) = f_{n+1}(x)$. Of course, the Ward identity 
allows one to reduce the whole tower to $f_0(x)$ and $f_1(x)$ and the action of the differential to the 
$2 \times 2$ connection above. 

Finally, in order to package the Ward identity in a better format, we can think about it as the differential on a complex.
The complex consists of holomorphic differential forms on the complex plane parameterized by $z$. 
The differential is 
\begin{equation}
d: \omega \to \partial^{(z)} \omega + (z^2 + x) \omega
\end{equation}
where $\partial^{(z)}$ acts on the $z$ direction only. Forms closed under the differential can be used 
in contours integrals of the form 
\begin{equation}
f_\omega(x) = \oint \omega e^{\frac{z^3}{3} + x z}  dz
\end{equation}
and exact forms integrate to zero. 
The differential commutes with the holomorphic connection 
\begin{equation}
\partial_x f_{\omega}(x) = f_{\partial_x \omega + z \omega}(x) 
\end{equation}

The description of the Airy function D-module as an infinite-dimensional complex may seem rather redundant 
compared to the simple $2 \times 2$ holomorphic connection we started with. It may be better suited, though, 
if one needs to describe the D-module as an object in some derived category of D-modules. 

This may appear to be only a matter of mathematical formalization, but it is likely to 
become a bit more physical if we want to use this setup in order to define a complex Lagrangian (BAA) 
brane in $T^* \C \simeq \C^2$ supported on $p^2 + x =0$ as a deformation of elementary BAA branes of the form 
$p=z$. 

Turning $z$ into a dynamical 1d chiral multiplet with a $\frac{z^3}{3}$ boundary superpotential 
defines a BAA boundary condition supported $p^2 + x =0$ which we expect to precisely correspond to 
the above D-module. The infinite-dimensional complex is simply the Chan-Paton bundle defined by these auxiliary 
1d degrees of freedom. 

As the support of the D-brane is smooth, we could of course directly define the corresponding $p^2 + x =0$ BPS boundary condition 
in the $(4,4)$ sigma model with target $\C^2$. Converting the physical boundary condition into the data of a D-module, 
would then require extra work, such as computing the space of A-type morphisms from the brane to the 
elementary BAA branes supported on the constant $x$ fibers of $T^* \C$ and the parallel transport 
along the space of fibers. 
 
In order to appreciate better the relative usefulness of different D-module descriptions, we can look at more singular examples. 
It is convenient to include a formal $\hbar$ quantization variable in our formulae to help with semi-classical limits. 

We will look at D-modules modelled on general integrals of the form 
\begin{equation}
\oint \omega(z,x) e^{\frac{W(x,z}{\hbar}} 
\end{equation}
with $\omega(z)$ being a holomorphic form on some auxiliary space $\CZ$ parameterized by $z$. 

The differential on the complex of holomorphic forms on $\CZ$ will be 
\begin{equation}
d: \omega \to \hbar \partial^{\CZ} \omega+ \partial^{\CZ} W \wedge \omega
\end{equation}
and the holomorphic connection 
\begin{equation}
\hat p =  \hbar \partial_x \omega+ \partial_x W \omega
\end{equation}
It is useful to shift the degrees of forms so that the top form has degree $0$. 

Recall that the brane wrapping $p=0$ is described by the trivial D-module consisting of polynomials of $x$ 
acted upon by $\hat p = \hbar \partial_x$. The brane wrapping $x=0$ is described by the Fourier transform of that, 
modelled on the contour integral 
\begin{equation}
\delta(x) = \oint e^{x z}  dz
\end{equation}
Here the complex is generated by the forms $x^n z^m$ and $x^n z^m dz$, 
with differential 
\begin{equation}
x^n z^m \xrightarrow{d} \left(\hbar m x^n z^{m-1} + x^{n+1} z^m\right)dz
\end{equation}
The cohomology consists of vectors of the form $z^n dz$ with multiplication by $x$ acting as $- \hbar \partial_z$ and 
$\hat p$ acting as multiplication by $z$, as expected. 

A simple but non-trivial example is modelled on the Gaussian integral 
\begin{equation}
\oint \omega(z,x) e^{\frac{1}{\hbar} x z^2} 
\end{equation}
The complex is generated by the forms $x^n z^m$ and $x^n z^m dz$, 
with differential 
\begin{equation}
x^n z^m \xrightarrow{d} \left(\hbar m x^n z^{m-1} + 2 x^{n+1} z^{m+1}\right)dz
\end{equation}
and action (with or without an overall $dz$) 
\begin{equation}
\hat p(x^n z^m) = \hbar n x^{n-1} z^m + x^{n} z^{m+2}
\end{equation}

The cohomology consists of vectors of the form $x^n dz$ and vectors of the form $z^m dz$.
The multiplication by $x$ acts as 
\begin{equation}
\hat x (x^n dz) = x^{n+1} dz \qquad \hat x (z^m dz) = - \frac{\hbar}{2} (m-1) z^{m-2} dz
\end{equation}
and $\hat p$ acts as 
\begin{equation}
\hat p (x^n dz) =  \hbar (n-\frac12) x^{n-1} dz \qquad \hat p (z^m dz) = z^{m+2} dz
\end{equation}
We thus find the direct sum of two modules for the $\hat x$, $\hat p$ Heisemberg algebra: 
one consisting of vectors of the form $x^n dz$ and $z^{2n} dz$ and one consisting 
of the vectors of the form $z^{2n+1} dz$.
The latter summand represents a copy of the trivial brane wrapping $x=0$. This is evident under the change of variables $z^2 \to z$.
The former summand represents a single brane supported classically on $x p =0$, distinct from the simple sum of the two components. 

If we were to avoid looking too closely at $x=0$, we may describe the D-module as a 
meromorphic connection with a regular singularity at the origin, something like 
$\hbar \left( \partial_x + \frac{1}{2 x} \right)$. This would hide a whole extra brane 
sitting at $x=0$!

A simple way to understand the existence of the extra component is to observe that the measure allows for a change of variables
\begin{equation}
\oint z^{2 n+1} x^m e^{\frac{1}{\hbar} x z^2} = \frac{1}{2} \oint d(z^2) (z^2)^n x^m e^{\frac{1}{\hbar} x (z^2)}
\end{equation}
making the identification with a $x=0$ brane obvious. 

The system admits an interesting deformation, modelled on 
\begin{equation}
\oint \omega(z,x) e^{\frac{1}{\hbar} \left(x z^2 + 2 a z\right)} 
\end{equation}
This integral maps $dz$ to $x^{- \frac12} e^{- \frac{a^2}{\hbar x}}$ which suggests a single smooth brane 
supported on $p = \frac{a^2}{x^2}$. The deformation makes it quite clear that in the $a \to 0$ 
limit the brane wraps twice the $x=0$ plane and once the $p=0$ plane. 

Next, consider the following example with two auxiliary fields, modelled on 
\begin{equation}
\oint \omega(u,v,x) e^{\frac{1}{\hbar}x u v} 
\end{equation}
Notice immediately that the system has an additional $U(1)_{uv}$ symmetry rotating 
$u$ and $v$ in opposite directions. The symmetry implies that the cohomology 
is $U(1)_{uv}$-invariant, as the Lie derivative of a form along the $U(1)_{uv}$ vector field
equals the anti-commutator of $d$ with the operation of contraction 
with the $U(1)_{uv}$ vector field. Concretely, the integrand must preserve the 
$U(1)_{uv}$ for the integral to be non-zero. 

The cohomology in degree $0$ is intuitive: it consists of forms $x^n du dv$
and $(uv)^n du dv$.  
The multiplication by $x$ acts as 
\begin{equation}
\hat x (x^n du dv) = x^{n+1} du dv \qquad \hat x (u^n v^n du dv) =  \hbar n (uv)^{n-1} du dv
\end{equation}
and $\hat p$ acts as 
\begin{equation}
\hat p (x^n du dv) =  \hbar \left(n-1 \right) x^{n-1} du dv \qquad \hat p  (u^n v^n du dv) = (uv)^{n+1} du dv
\end{equation}
This module is a non-trivial extension of the basic $x=0$ and $p=0$ modules: the forms with positive powers of $x$ form a sub-module,
but $du dv$ is mapped to $x du dv$ by multiplication by $x$. 

It is important to observe that there is also cohomology in degree $-1$, generated by $u dv + v du$. 
It corresponds to the integral 
\begin{equation}
\oint d (uv) e^{\frac{1}{\hbar}x u v} 
\end{equation}
which is related to an $x=0$ brane by the obvious change of variable. Indeed, the module consists of forms $(uv)^n d(uv)$ with $\hat p$ acting as multiplication by $uv$ and multiplication by $x$
as derivative by $uv$. Thus the full system involves two branes, one in degree $0$ and one in degree $-1$. Although the two branes live in different degrees, 
they can communicate by morphisms of degree $1$, i.e. extensions. 

Next, we can consider a system involving $N$ auxiliary variables $z_i$, 
modelled on 
\begin{equation}
\oint \omega(z,x) e^{\frac{1}{2\hbar} \sum_i (x-a_i) z_i^2} 
\end{equation}
If we assume that the $a_i$ constants are all different, the cohomology appears to consist of $N+1$ separate modules. 
One module is generated by the top form $\prod_i dz_i$. It is a complicated module, supported on $p \prod_i (x-a_i) = 0$. 
The other modules are generated by $z_k \prod_i dz_i$ and are isomorphic to the basic module associated to $x = a_i$. 

This system is a particular case of the general 
\begin{equation}
\oint \omega(z,x) e^{\frac{1}{2\hbar} z^t \left( x - A \right) z}
\end{equation}
where the $N \times N$ constant symmetric matrix $A$ is taken to have $N$ distinct eigenvalues. 

This general model, perhaps re-written as 
\begin{equation}
\oint \omega(z,x) e^{\frac{1}{2\hbar} z^t M(x) z}
\end{equation}
for some $N \times N$ matrix $M(x)$ which depends linearly on $x$,
is a reasonable toy model to describe the D-modules one can encounter 
along one-dimensional slices of a parameter space. It shows how one can get 
interesting extra cohomology supported on the locus where $M(x)$ has zeromodes. 

In the $U(1)$-invariant version of this problem, i.e.   
\begin{equation}
\oint \omega(u,v,x) e^{\frac{1}{2\hbar} u^t M(x) v}
\end{equation}
we can model the effect of an action which has $k$ zeromodes for generic values of $x$,
simply by taking $M$ to be an $N \times N+k$ matrix. 
Notice that the $U(1)$ symmetry requires the integrand to have $k$ more $u$'s than $v$'s

\subsection{An extra examples in $T^* \C^2$.}
Consider the D-module on $\C^2$ modelled on 
\begin{equation}
\oint \omega(u,v,w,x,y) e^{\frac{1}{\hbar}\left(x u v + y u w \right)} 
\end{equation}
with $x,y$ coordinates on $\C^2$ in $T^* \C^2$.

This is a toy model for a situation where generically the path integral has a bosonic zeromode of charge $1$
(here $x v + y w$) but at a special co-dimension $2$ locus has $2$ zeromodes of charge $1$ and one of charge $-1$. 

Because of the generic zeromode, $du dv dw$ is not a good measure of integration: 
\begin{equation}
du dv dw e^{\frac{1}{\hbar}\left(x u v + y u w \right)} = d \left( (-u dv dw - v du dw + w du dv)e^{\frac{1}{\hbar}\left(x u v + y u w \right)} \right) 
\end{equation}

Instead, we can find non-trivial cohomology in degree $-1$: 
\begin{equation}
\oint d(uv) d(uw) e^{\frac{1}{\hbar}\left(x u v + y u w \right)} 
\end{equation}
simply represents the standard brane at $x=y=0$. 

This example has obvious higher dimensional generalizations. 
\subsection{Matrix examples}
It is instructive to look at a more general family of problems with bosonic zeromodes.

\subsection{Symplectic bosons conformal blocks}
The simple finite-dimensional examples we considered in the previous section are actually quite close to 
our main subject of interest: conformal blocks for symplectic bosons transforming in a representation $M$ 
of some group $G$, coupled to some $G$-bundle $E$ with associated bundle $M_E$. 

The path integral is holomorphic, with action $\langle Z, \bar \partial_E Z \rangle$. General correlation functions can be written as 
\begin{equation}
\oint \omega[Z] e^{\langle Z, \bar \partial_E Z \rangle}
\end{equation}
with $\omega[Z]$ including the path integral measure and the fields inserted in the correlation function. 

In analogy to the finite-dimensional case, one can consider a D-module on $\Bun(G,C)$
built as a complex of possible $\omega[Z]$, with a differential
\begin{equation}
d: \omega \to \hbar \partial^{Z} \omega+ \int_C \langle \delta Z, \bar \partial_E Z \rangle \wedge \omega
\end{equation} 
which guarantees that the overall path integral is at least formally a closed form and 
Ward identities are satisfied. 

In other words, the cohomology of the complex is what we would usually call a conformal block: 
a collection of correlation functions which solve the Ward identities. As in the finite-dimensional examples, 
the space of conformal blocks will jump in complicated ways at loci where the number of zeromodes 
of the $\bar \partial_E$ operator changes. 

Because of that, rather than passing to the cohomology it is natural to define formally the space of conformal blocks and its D-module structure 
 in a derived sense as the whole complex, at least formally. 

In order to give a somewhat less formal definition, one could seek some sort of intermediate 
description, which integrates out most of the degrees of freedom in the $Z$ fields 
but still leaves a finite-dimensional path integral undone and describes conformal blocks 
as a non-trivial complex which does not jump wildly as we move along $\Bun(G,C)$.  

A suggestion may come from the standard sewing construction of Riemann surfaces 
and conformal blocks. In a gauge where the symplectic bosons are single-valued around the 
sewing fixtures, the conformal block is assembled from $n$-point functions of 
vacuum descendants on spheres. The latter are fully and easily determined 
by the Ward identities. The data of the bundle and all the subtleties
which concern us arise in the gluing operation. 

The whole data of the three-point functions can be expressed as a Gaussian functional of the 
Fourier modes $Z^{(i)}_{- n - \frac12}$ of the symplectic bosons at the punctures. The gluing maps are also encoded 
into Gaussian functionals of the modes at the pairs of punctures being sewn together. Thus the whole conformal block 
becomes a Gaussian integral over the modes of the symplectic bosons at the punctures being sewn together. 

This Gaussian integral is still infinite-dimensional, but it is simpler than the original path integral: effectively, we have 
integrated out all modes which are not holomorphic away from the sewing fixtures and we are only left with the 
work of imposing the gluing constraints. The D-module of conformal block should be well represented by this reduced Gaussian 
integral. 

In order to simplify the problem further, we could try to integrate out non-holomorphic modes everywhere except than at a few selected points on the Riemann surface. 
That means adding sources to the equations of motion at these points and then requiring these sources to vanish by adding Lagrange multipliers.

Alternatively, it should be possible to give local descriptions in the neighbourhood of a point in $\Bun(G,C)$ by computing the 
action at nearby points for the modes which are zeromodes at the point and modelling the conformal blocks on the resulting zeromode path integral.

\bibliography{3dTwists}{}
\bibliographystyle{JHEP}
\end{document}